\documentclass[iop]{emulateapj}

\usepackage{amsmath,amssymb, graphicx, natbib, multirow, morefloats}
\bibstyle{apj}

\newcommand{\metal}{\mathrm{[Fe/H]}}
\newcommand{\um}{$\mu$m}
\newcommand{\youthpercent}{$\sim$3\%}

\newcommand{\initialcut}{2,665}
\newcommand{\finalcut}{300}
\newcommand{\totalsample}{175}
\newcommand{\wthreesample}{175}
\newcommand{\wfoursample}{13}
\newcommand{\orionsample}{seven}
\newcommand{\minusorion}{168}
\newcommand{\lithiumstars}{four}
\newcommand{\minitab}[2][l]{\begin{tabular}{#1}#2\end{tabular}}

\slugcomment{Accepted for publication in \apj}
\shorttitle{M Dwarfs with \textit{WISE} Excesses}
\shortauthors{Theissen \& West}

\begin{document}

\title{Warm Dust around Cool Stars: Field M Dwarfs with \textit{WISE} 12 or 22 Micron Excess Emission}

\author{Christopher A. Theissen and Andrew A. West}
\affil{Department of Astronomy, Boston University, 725 Commonwealth Avenue, Boston, MA 02215, USA}

\email{ctheisse@bu.edu}

\begin{abstract}
Using the SDSS DR7 spectroscopic catalog, we searched the \textit{WISE} AllWISE catalog to investigate the occurrence of warm dust, as inferred from IR excesses, around field M dwarfs (dMs). 
We developed SDSS/\textit{WISE} color selection criteria to identify 175 dMs (from 70,841) that show IR flux greater than typical dM photosphere levels at 12 and/or 22 \um, including seven new stars within the Orion OB1 footprint.
We characterize the dust populations inferred from each IR excess, and investigate the possibility that these excesses could arise from ultracool binary companions by modeling combined SEDs.
Our observed IR fluxes are greater than levels expected from ultracool companions ($>3\sigma$).
We also estimate that the probability the observed IR excesses are due to chance alignments with extragalactic sources is $<$ 0.1\%. 
Using SDSS spectra we measure surface gravity dependent features (K, Na, and CaH 3), and find $<$ 15\% of our sample indicate low surface gravities. 
Examining tracers of youth (H$\alpha$, UV fluxes, and Li absorption), we find $<$ 3\% of our sample appear young, indicating we are observing a population of field stars $\gtrsim$ 1 Gyr, likely harboring circumstellar material. 
We investigate age-dependent properties probed by this sample, studying the disk fraction as a function of Galactic height. 
The fraction remains small and constant to $|Z| \sim 700$ pc, and then drops, indicating little to no trend with age. 
Possible explanations for disks around field dMs include: 1) collisions of planetary bodies, 2) tidal disruption of planetary bodies, or 3) failed planet formation.
\end{abstract}


\section{Introduction}

	Disks are an important part of stellar and planetary evolution, providing the building blocks for planetary systems around all types of stars. Since the discovery of a disk-like structure encircling the A0V star Vega \citep{aumann:1984:l23}, the study of circumstellar disks has expanded our understanding of stellar and planetary evolution. After its launch in 1983, the \textit{Infrared Astronomical Satellite} \citep[\textit{IRAS;}][]{neugebauer:1984:l1} observed an unprecedented number of young stellar objects (YSOs), in addition to more evolved stars, akin to Vega, which were identified by their excess infrared (IR) fluxes \citep{zuckerman:2001:549}. For the first time, a diverse range of disks were characterized by their IR signatures \citep{lada:1987:1}, from pre-main sequence stars showing near-IR (NIR) excesses due to a hot shell of infalling material, to colder disks analogous to the Kuiper belt, showing far-IR (FIR) excess fluxes \citep[e.g., Vega;][]{su:2005:487}. The \textit{Spitzer Space Telescope} \citep{werner:2004:1}, launched in 2003, allowed for further characterization of disk populations, categorizing the transitional disk phase \citep{espaillat:2014:}, where IR gaps appear as signposts of planetary construction. The high sensitivity and long wavelength range (from $10^{-4}$ mJy at 3.6 \um\ to 10 mJy at 160 \um) of \textit{Spitzer} allowed for some of the most detailed studies of nearby star-forming regions to date \citep[][and references therein]{williams:2011:67}. 
	
	As a successor to \textit{Spitzer}, \textit{Herschel} \citep{pilbratt:2010:l1} was able to probe even colder disk populations, with spectral energy distributions (SEDs) peaking between 55--670 \um, and at an unprecedented resolution. Studies using \textit{Herschel} have investigated the properties of cold debris disks \citep{eiroa:2010:l131,greaves:2010:l44,acke:2012:a125,eiroa:2013:a11}. \textit{Herschel} was also able to measure a new class of cold debris disks, with temperatures of $\sim$22 K \citep{eiroa:2011:l4,ertel:2012:a148,krivov:2013:32}. Although there is still some debate on whether these observed FIR excesses originate from cold circumstellar material versus Galactic contamination \citep{gaspar:2014:33}, their existence demonstrates the sensitivity of \textit{Herschel}. 
	
	While \textit{Spitzer} and \textit{Herschel} provided targeted IR observations, much of the sky remained void of IR observations. In 2011, The \textit{Wide-field Infrared Survey Explorer} \citep[\textit{WISE};][]{wright:2010:1868} completed an all-sky IR survey, providing the ability to study the IR properties of field stars that are not members of stellar associations or young star-forming regions. The \textit{WISE} observational strategy allows for a ``blind" search for stars exhibiting IR excesses above their photospheres across the sky. \textit{WISE} observations are currently helping to answer stellar and planetary evolution questions regarding the mechanisms responsible for disk dispersal, the detailed timescales on which they act, and how they differ with stellar mass \citep[e.g.,][]{avenhaus:2012:a105,heng:2013:}. 
		
	M dwarfs, being the most populous \citep[$\sim$70\% of the stellar population;][]{bochanski:2010:2679} and longest living stellar constituents \citep[main sequence lifetimes $\sim$$10^{12}$ yrs;][]{laughlin:1997:420}, offer a laboratory in which to study the statistical properties of our Galaxy. In addition, recent studies have shown that M dwarfs have high probabilities of harboring Earth-sized terrestrial planets \citep{howard:2012:15,swift:2012:59,dressing:2013:95}, allowing studies to constrain the number of planetary systems within the Milky Way. However, despite huge advances in the study of circumstellar material over the past decade, the detailed process of planetary formation around low-mass stars ($\lesssim 0.6M_\odot$) still elicits questions regarding the mass-dependence of circumstellar disk evolution.
	
	The earliest investigations of disks around low-mass stars used \textit{IRAS} to detect signatures of cold dust ($< 100$ K) by identifying small 25 \um\ excesses around stars in young star-forming regions. Many of these detections were found to be false-positives \citep{zuckerman:2001:549, song:2002:514, plavchan:2005:1161}, possibly due to the low spatial resolution of \textit{IRAS} and consequent blending of background sources with target sources, or a detection threshold bias, leaving a dearth of low-mass stars with cold disks. One such study undertaken by \citet{plavchan:2005:1161} investigated the paucity of cold disks around M dwarfs, using small mid-IR (MIR) excesses ($> 10$ \um) as a proxy, and postulated that rapid planet formation \citep{weinberger:2004:2246,boss:2006:501} may be a cause for the dearth of cold disks around low-mass stars older than 10 Myr. However, recent studies have suggested that this deficiency is likely due to \textit{IRAS} detection limits \citep{forbrich:2008:1107, plavchan:2009:1068}. 
	
	\citet{forbrich:2008:1107} used the \textit{Spitzer} Multiband Imaging Photometer (MIPS; 24, 70, and 160 \um\ bands) to study NGC 2547, a cluster with an age of 30--40 Myr, and specifically targeted low-mass stars within the region. Using the FIR abilities of \textit{Spitzer}, \citet{forbrich:2008:1107} were able to increase the number of known M dwarfs with cold disks from six to fifteen, as inferred from their 24 \um\ excesses. In addition, \citet{plavchan:2009:1068} completed \textit{Spitzer} observations of 70 main sequence stars between ages of 8 Myr to 1.1 Gyr and found that small 24 \um\ excesses ($\sim$10\% above the expected photospheric flux), associated with cold disks, were common around young stars of all spectral types. \citet{plavchan:2009:1068} also observed that large 70 \um\ excesses ($\gg 100\%$ above the expected photospheric flux) were more common around stars with $T_{\ast} >$ 5,000 K in the 200--500 Myr range than stars with $T_{\ast} <$ 5,000 K in the 8--50 Myr range. The discrepancy in 70 \um\ excesses may be due to higher-mass stars having larger, brighter photospheres, irradiating a larger portion of the disk. It is also probable that disks around higher mass stars are more massive due to the larger protostellar envelope from which they are formed \citep{plavchan:2009:1068}. Another important consideration is observations at a specific wavelength probe different sections of the disk for different stellar temperatures \citep{kennedy:2009:1210}. The number of M dwarfs showing signs of dust is currently insufficient to properly address the questions of disk evolution and structure for the lowest mass stars.
	
	The excess differences between hot and cool main sequence stars found by \citet{plavchan:2009:1068} can be explained by invoking mass-dependent evolution of circumstellar disks. Several studies have found that young star-forming regions ($< 5$ Myr) show drastically different fractions of stars exhibiting IR excesses above their photospheres as a function of spectral type \citep{carpenter:2006:l49, lada:2006:1574, luhman:2008:1375, carpenter:2009:1646, luhman:2010:111}. Even regions of approximately similar ages showed significant differences in their disk fractions, possibly as a result of the density of stars within the region \citep[e.g., open versus globular clusters;][]{luhman:2008:1375, luhman:2010:111}. Other studies have pointed to metallicity playing a significant role in circumstellar disk evolution \citep{yasui:2009:54,yasui:2010:l113}. The planet-metallicity correlation \citep{gonzalez:1997:403,laws:2003:2664,fischer:2005:1102,johnson:2010:905,rojas-ayala:2010:l113} may imply that rapid formation of planets in these higher metallicity systems, could disperse the disk on faster timescales. These effects may be significant for M dwarfs, which are prone to low-mass planet formation \citep{swift:2012:59,dressing:2013:95}. However, other studies have suggested that stellar metallicity does not correlate with the occurrence of Earth-sized terrestrial planets \citep{mayor:2011:,buchhave:2012:375}. Observations indicate that Earth-sized planets appear to be more common around M dwarfs than Jupiter-sized planets, suggestive that metallicity may not play a significant role in planet formation for the lowest-mass stars \citep{dressing:2013:95}. The link between metallicity, planet formation, and disk evolution around low-mass stars still requires further investigation. 
	
	In recent years, a new class of older stars ($\gtrsim 1$ Gyr) exhibiting IR excess has been identified \citep[][]{rhee:2008:777,moor:2009:l25,fujiwara:2010:l152,melis:2010:l57,weinberger:2011:72}. These solar-type stars (FGK-spectral types) pose questions regarding our current understanding of disk evolution. One such star, BD +20 307, investigated by \citet{weinberger:2011:72} was found to have a large IR excess ($L_{IR} / L_\ast\approx 10^{-2}$), speculated to have been caused by the collision of two planetary-sized bodies within the terrestrial zone. Such a phenomenon has not been observed around M dwarfs, presumably due to their intrinsic faintness coupled with the sensitivity limits of previous IR surveys (e.g., \textit{IRAS}). However, due to the extremely long lifetimes of M dwarfs ($\sim$$10^{12}$ yrs), these stars give us the ability to study the long-term stability of planetary systems, and other possible evolutionary scenarios such as $2^\mathrm{nd}$ generation planet formation \citep{melis:2009:1964,melis:2010:470,perets:2010:}. 
	
	One strategy to increase source detections, and hence statistics, is to increase the solid angle of sky surveyed and the spectral coverage of observations. With the advent of large-area and all-sky surveys, the capability to study statistically significant populations of field stars has recently become available. At optical wavelengths, the Sloan Digital Sky Survey \citep[SDSS;][]{york:2000:1579} has become invaluable in creating large catalogs of low-mass stars \citep{bochanski:2010:2679,west:2011:97}. In the NIR, the Two Micron All-Sky Survey \citep[2MASS;][]{skrutskie:2006:1163} has been extremely useful in discovering hot dust populations ($>$ 1,000 K), making up the ``pre-transitional" disk phase of protostellar evolution \citep{espaillat:2014:}. The \textit{WISE} mission recently completed an all-sky survey in the MIR. The all-sky nature and depth of \textit{WISE} has made it possible to search for stars that exhibit infrared excesses and are not in young star-forming regions \citep[e.g.,][]{avenhaus:2012:a105, kennedy:2012:91}. Together, SDSS, 2MASS, and now \textit{WISE} provide photometric coverage over an extremely large solid-angle of sky, from the optical to MIR. These surveys have given astronomers the ability to search for low-mass objects in young star-forming regions \citep{dawson:2012:310} and nearby young moving groups \citep[NYMGs;][]{shkolnik:2011:6}, study the statistical properties of our Galaxy \citep{bochanski:2008:,bochanski:2010:2679,west:2008:785,west:2011:97}, and explore the optical and IR properties of vast numbers of sub-solar objects \citep{hawley:2002:3409, kirkpatrick:2011:19}. 2MASS and \textit{WISE} are useful in constraining the effective temperature for low-mass stars, since their SEDs peak and turn over in the NIR.
	
	A difficulty in searching for new, young, low-mass stars is the uncertainty in determining stellar ages. Many different methods are used to estimate ages for stars, including: 1) determining the lithium depletion boundary \citep{cargile:2010:l111}; 2) using spectroscopic tracers such as sodium \citep{schlieder:2012:114} and H$\alpha$ \citep{west:2008:785,west:2011:97}; 3) rotation and activity \citep{mohanty:2003:451, reiners:2008:1390}; and 4) ultraviolet flux \citep{shkolnik:2011:6, schlieder:2012:80}. Each of these youth indicators has its caveats, including the fact that metallicity may affect lithium equivalent width measurements \citep{pinsonneault:1997:557}, or binarity may affect rotation and activity \citep{morgan:2012:93}. As such, it is important to test for multiple tracers of youth to definitively classify a star as young ($\lesssim 100$ Myr) versus old ($\gtrsim 1$ Gyr). Age is an important parameter for stars exhibiting IR excess as it potentially helps to differentiate stars harboring primordial dust from $2^\mathrm{nd}$ generation scenarios that create circumstellar material.
	
	A limitation in the search for disks around low-mass stars has been the lack of a reliable input catalog. Widely used stellar catalogs have been biased towards bluer stars, due to the intrinsic faintness of M dwarfs and blue bias of previous filters (e.g., \textit{Hipparcos}, Tycho-2, etc.) This study aims to increase the number of known M dwarfs with MIR excesses by combining data from SDSS, 2MASS, and \textit{WISE}. We searched the \textit{WISE} AllWISE source catalog for signs of IR excess around the M dwarf spectroscopic sample of \citet[70,841 M dwarfs;][]{west:2011:97}. In \S\ref{data} we outline our selection criteria for high-probability disk candidates from the sample compiled by \citet{west:2011:97}, define SDSS and \textit{WISE} color-color criteria for selecting M dwarfs with IR excesses, and address the issue of interstellar reddening. Combining SDSS, 2MASS, and \textit{WISE} photometry, we create SEDs for our candidates in \S\ref{methods}, and explore the possibility that our IR excesses could be due to an ultracool companion, Galactic IR cirrus, or extragalactic contamination. In \S\ref{results} we characterize our stars and observed dust content by: 1) modeling the dust populations using idealized parameters (\S\ref{dust}), and 2) investigating a number of tracers for activity, surface gravity, and accretion as proxies for youth (\S\ref{youth}). We also briefly discuss stars of interest at the end of \S\ref{results}. A discussion of our sample and possible interpretations and scenarios for circumstellar material around field dwarfs follows in \S\ref{discussion}. Lastly, in \S\ref{summary} we summarize our findings.

\section{Data}\label{data}

	To construct SEDs and identify stars with excess IR flux requires data over a large wavelength regime, from the optical to the MIR. Additionally, spectroscopic data can be used to determine stellar parameters (\S\ref{methods}) and youth diagnostics (\S\ref{youth}). To obtain data over the wavelength coverage needed, we combined data sets from SDSS, 2MASS, and \textit{WISE}.
	
	SDSS has played an integral role in the statistical study of large stellar populations, specifically low-mass stars. The magnitude limits of SDSS allow the survey to probe M dwarfs out to distances $>$ 1,000 pc, yielding vast photometric catalogs of M dwarfs \citep[$> 30$ million;][]{bochanski:2010:2679}. In addition, the medium resolution ($R \approx$ 1,800) spectroscopic pipeline for SDSS has produced prodigious spectroscopic samples of M dwarfs \citep[][hereafter W08, W11, respectively]{west:2008:785,west:2011:97}. W11 used SDSS Data Release 7 \citep[DR7;][]{abazajian:2009:543} to compile a spectroscopic catalog of 70,841 M dwarfs. All of the stars were visually spectral typed using the Hammer IDL routine \citep{covey:2007:2398}. We also measured radial velocities (RVs) using a cross-correlation-like program, a Python version of the \texttt{xcorl.pro} IDL procedure \citep{mohanty:2003:451,west:2009:1283}, comparing each spectrum with the appropriate M dwarf template \citep{bochanski:2007:531}. This method yields typical uncertainties of $\sim$7 km s$^{-1}$ \citep{bochanski:2007:531}. 
		
	NIR and MIR data were obtained from the 2MASS point source catalog \citep{skrutskie:2006:1163} and \textit{WISE} AllWISE source catalog \citep{wright:2010:1868,mainzer:2011:53}, respectively. 2MASS provides full-sky coverage in three NIR bands ($J$: 1.25, $H$: 1.65, and $K_s$: 2.17 \um). The SEDs of M dwarfs peak in the NIR, thus, 2MASS photometry is especially critical in constraining the stellar effective temperature. In addition, 2MASS provides quality flags, useful to W11 in building the spectroscopic catalog and to this study, for ensuring point-source (\textsc{gal\_contam}), high-quality photometry (\textsc{rd\_flg} \& \textsc{cc\_flag}). We required these quality flags for our own cuts outlined below. 
			
	In 2009, NASA launched \textit{WISE} with the mission of completing an all-sky survey in four MIR bands \citep{wright:2010:1868}. These bands were centered at 3.4, 4.6, 12, and 22 \um, hereafter referred to as $W1$, $W2$, $W3$, and $W4$, respectively. \textit{WISE} is the most sensitive all-sky MIR survey to date, surpassing \textit{IRAS}, and the recent survey completed by \textit{AKARI} \citep{murakami:2007:369} in 2007. $W1$, $W2$, $W3$, and $W4$ have angular resolutions of 6.1\arcsec, 6.4\arcsec, 6.5\arcsec, and 12\arcsec, and 5$\sigma$ point source sensitivities of 0.068, 0.098, 0.86 and 5.4 mJy, respectively. \textit{WISE} bands, specifically the 12 and 22 \um, probe the wavelength regime where warm ($T\sim30$--300 K) dust populations peak in the observed SED. The \textit{WISE} photometric pipeline also provides a number of quality flags useful in selecting point sources including the contamination and confusion flag (\textsc{cc\_flag}), and extended source flag (\textsc{ext\_flag}).

\subsection{Sample Selection}\label{sample}

	Starting with the W11 catalog of 70,841 spectroscopically confirmed M dwarfs, we selected a sample of stars showing IR excesses above their photospheres starting with the following selection process (cuts 4, 5, and 6 were taken from the \textit{WISE} Explanatory Supplement\footnote{http://wise2.ipac.caltech.edu/docs/release/allwise/expsup/sec2\_2a.html}; the number of stars left after each cut is given in parenthesis at the end of each selection criterion):
\begin{enumerate}
\item We matched the W11 catalog to the \textit{WISE} AllWISE source catalog\footnote{http://irsa.ipac.caltech.edu/Missions/wise.html}, matching only to unique \textit{WISE} counterparts within 5\arcsec. (66,890 stars)
\item The low angular resolution of \textit{WISE} (FWHM $= 6.1$\arcsec, 6.4\arcsec, 6.5\arcsec, and 12\arcsec\ in $W1$, $W2$, $W3$, and $W4$, respectively) required selection criteria to ensure that disk candidates were not contaminated by nearby sources. Using the SDSS-II CasJobs site\footnote{http://casjobs.sdss.org/CasJobs/} we obtained all DR7 primary photometric objects within 6\arcsec\ of our stars (to match the $W4$ beam) and kept only stars did not have another primary photometric object within a 6\arcsec\ radius. We applied the same criteria to stars with another 2MASS photometric object within 6\arcsec\ (\textsc{n\_2mass} $< 2$)\footnote{Definitions for \textit{WISE} flags can be found at http://wise2.ipac.caltech.edu/docs/release/allwise/expsup/sec2\_1a.html}. (57,544 stars)
\item We required high-quality photometry from the optical through the NIR. To achieve this, we kept only stars with reliable SDSS photometry as determined in W11 (\textsc{goodphot} = 1\footnote{Defined by SDSS photometric processing flags: SATURATED, PEAKCENTER, NOTCHECKED, PSF\_FLUX\_INTERP, INTERP\_CENTER, BAD\_COUNTS\_ERROR set to zero in the $r$, $i$, and $z$ bands.}), and photometry in all three 2MASS bands. (43,253 stars)
\item We kept only stars with source morphology that was consistent with a single point-spread function (PSF) in \textit{WISE} (\textsc{ext\_flag} $= 0$). (42,892 stars) 
\item Stars without a contamination and confusion flag in $W3$ (\textsc{cc\_flag} $= 0$) were kept. This removed sources that could be due to diffraction spikes or scattered light from a nearby bright source. We did not require this for $W1$ or $W2$ since these bands were not used for our SED fitting, and we were most concerned with real excesses at 12 \um. (42,632 stars)
\item Stars that had a signal-to-noise (S/N) ratio (SNR) $\geq3$ in $W1$, $W2$, or $W3$ were included. Due to the distances to stars in the W11 catalog (100--2,000 pc), we expect real excesses to be faint, therefore we chose this S/N to be as inclusive as possible. An IR excess may also be observed in $W4$, however, as this was the shallowest \textit{WISE} band, we did not require a minimum $W4$ S/N. We will discuss 22 \um\ excesses in our sample in \S\ref{identifying}. (2,698 stars) 
\item Although the M giant contamination rate for the W11 catalog was estimated to be less than 0.5\%, we removed potential dusty giants from our sample. Using the results from \citet{bessell:1988:1134} we assumed any star with a $J-H > 0.8$ has a significant probability of being an M giant. We kept stars: 1) that had $J-H \leq 0.8$; or 2) that had $J-H > 0.8$ and whose total USNO-B proper motions were greater than the total uncertainty ($\mu_\mathrm{tot} > \sigma_{\mu_\mathrm{tot}}$) and whose total proper motions were not zero ($\mu_\mathrm{tot} \ne 0$). (\initialcut\ stars) 
\end{enumerate}

	These initial selection criteria reduced the sample size to \initialcut\ stars with reliable photometry in SDSS through $W3$ bands and no source confusion or contamination. We investigated other possible explanations for IR excesses around field dwarfs, specifically an ultracool companion or galactic/extragalactic contamination sources. We explore the likelihood of our IR excesses being attributable to these interlopers in \S\ref{binarity} and \S\ref{contamination}, respectively.

\subsubsection{Identifying the Disk Candidates}\label{identifying}

	\citet[][hereafter A12]{avenhaus:2012:a105} used low-mass stars from the RECONS 100 nearest star systems\footnote{http://www.chara.gsu.edu/RECONS/TOP100.posted.htm} \citep{jao:2005:1954, henry:2006:2360} to search for MIR excesses around K and M dwarfs using \textit{WISE}. This nearby sample offered high S/N \textit{WISE} observations, as well as accurate parallax measurements and absolute magnitudes. Although A12 did not find any MIR excesses, presumably due to the older age of the nearby stellar sample ($\gtrsim 30$ Myrs), they were able to determine typical MIR colors for K and M dwarfs. The A12 excess criteria require absolute magnitudes as a proxy for stellar effective temperature, hence distances are required. Our sample has photometric distances, with typical uncertainties of $\sim$20\%, originally computed by W11 using the photometric parallax polynomials from \citet{bochanski:2010:2679}, and then dust-corrected by \citet{jones:2011:44}, we use the dust-corrected of these distances in our analysis.
	
\begin{figure*}
\centering
 \plotone{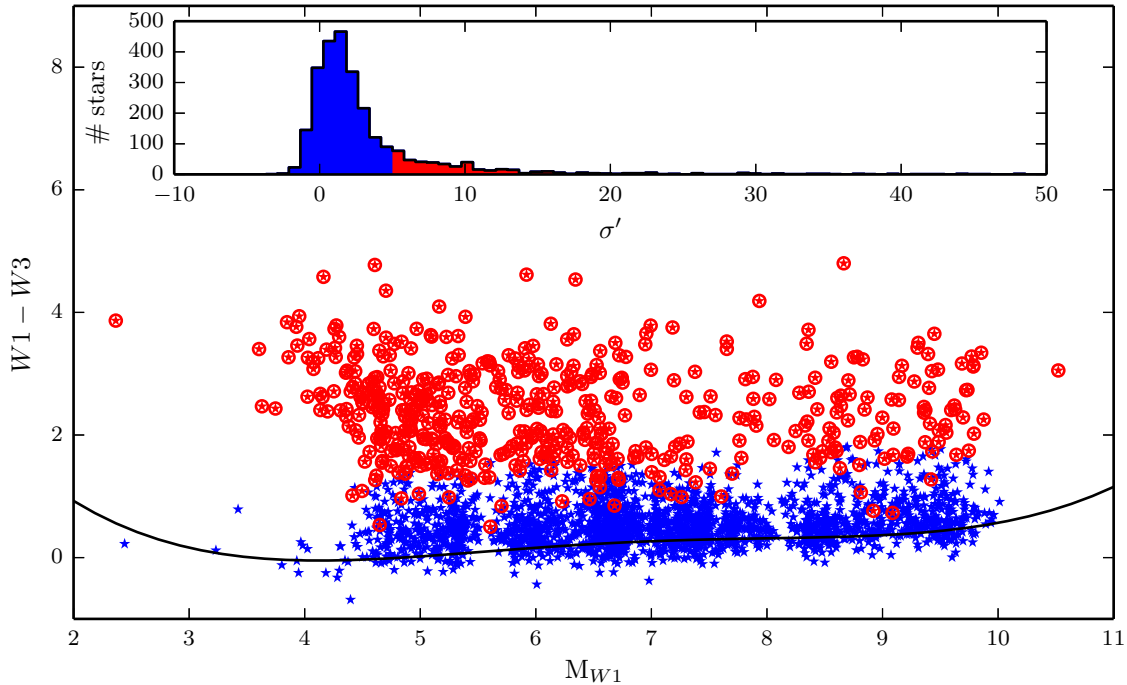}
\caption{\textit{WISE} Color-magnitude plot for the stars that made our initial cuts but prior to visual inspection (480 excess and 2,185 non-excess stars). The A12 polynomial is plotted (black line), along with the distribution of excess significance as determined by the A12 criteria. Stars $\ge 5\sigma'$ (defined in A12) are plotted as red stars with circles, representing the disk candidates. Stars below the $5\sigma'$ limit are plotted in blue stars. The overlap region is due to excess candidates ($\ge 5\sigma'$) with small \textit{WISE} measurement uncertainties, and non-candidates ($< 5\sigma'$) with large uncertainties. The inset plot shows the histogram of excess significances ($\sigma'$) as defined in A12.
\label{fig:W3excess}}
\end{figure*}

\begin{figure*}
 \plotone{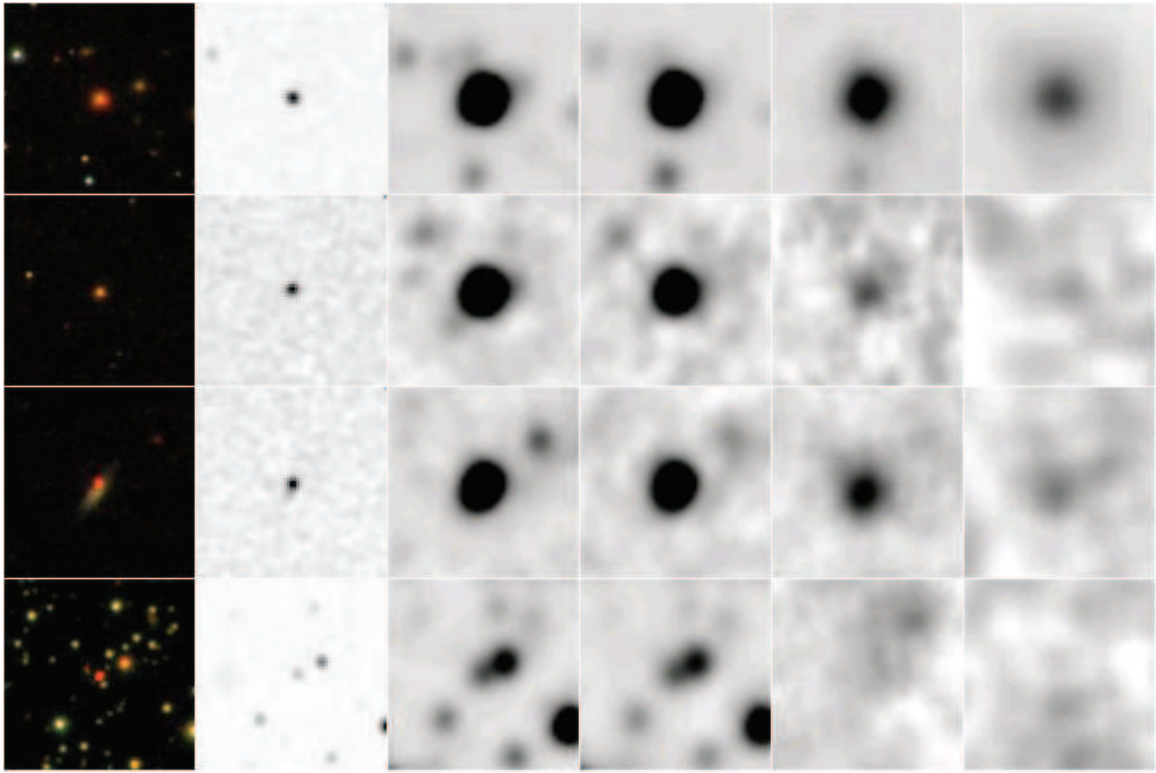}
\caption{Images in SDSS (\textit{gri} composite), 2MASS $J$, and \textit{WISE} (3.4, 4.6, 12, and 22 \um, respectively) centered on the 3.4 \um\ source; each image is 1\arcmin\ $\times$ 1\arcmin. Top row: Example of an object that was included in our combined sample with \textsc{quality} $= 1$. Second row: Example of an object that was included in our combined sample with \textsc{quality} $= 2$. Third row: Example of a star that was not included in our sample due to superposition of a galaxy in SDSS. Bottom row: Example of a star that was not included in our sample due to crowding.\label{fig:GoodBad}}
\end{figure*}

	A12 defined their metric of IR excess significance as 
$$
\sigma' = \frac{e}{\sigma}\frac{1}{\sqrt{\chi^2}},
$$
where $e$ is the deviation from the A12 polynomial, $\sigma$ is the \textit{WISE} measurement error, and $\chi^2$ is the overall goodness-of-fit for the A12 color-magnitude relationship. Using the M$_{W1}$-SpT excess criteria of A12, we found that 480 of the \initialcut\ stars from our initial cuts showed significant IR excess ($\ge 5\sigma'$) in their $W1-W3$ color, as shown in Figure~\ref{fig:W3excess}. Visual inspection of these 480 stars in SDSS $gri$ composite images\footnote{http://cas.sdss.org/dr7/en/tools/chart/list.asp} and spectra showed that 171 of the 480 candidate stars were contaminated by nearby galaxies or bright sources, or were misclassified, appearing spectroscopically as either galaxies or K dwarfs. This visual inspection left 309 remaining candidate stars. An example of stars kept versus stars that were removed due to contamination is shown in Figure~\ref{fig:GoodBad}.
	
\begin{figure}
 \plotone{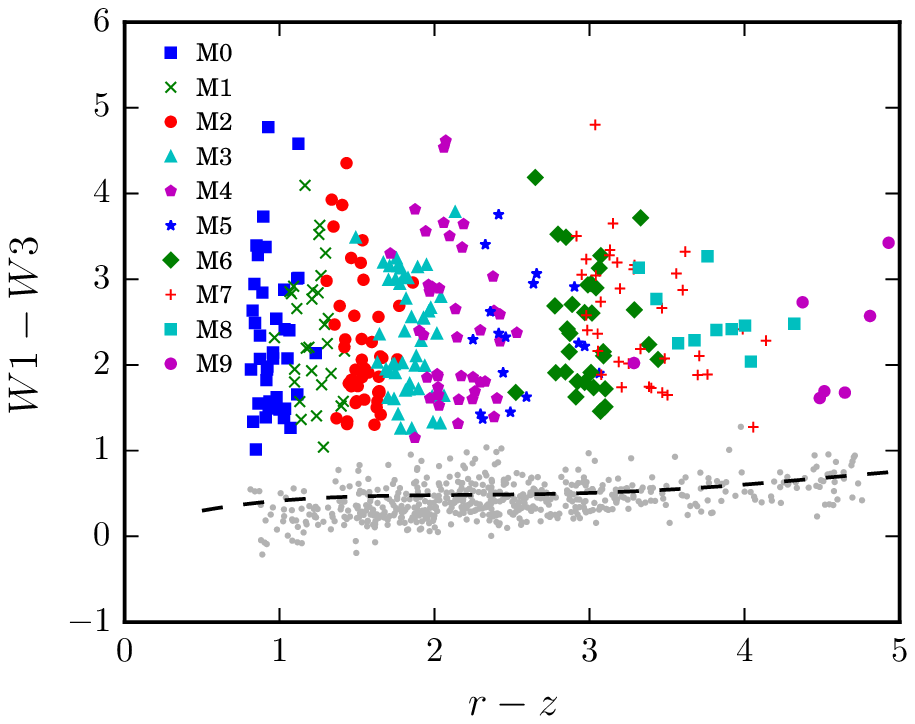}
 \plotone{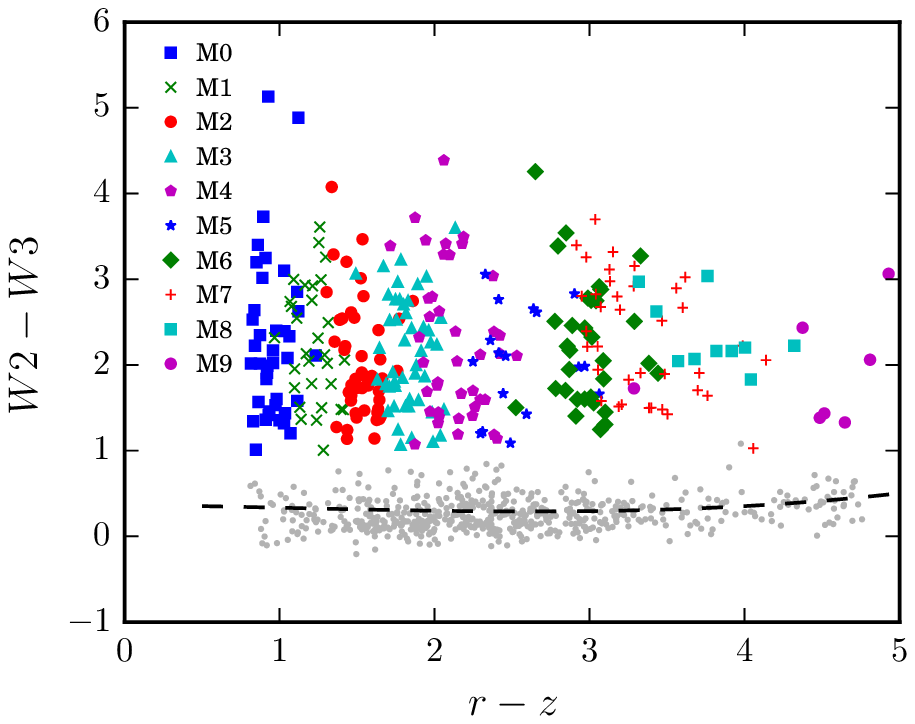}
\caption{Color-color diagrams for the 309 stars showing 12 \um\ excesses as defined by A12. Stars are separated by spectral type using the colors and symbols in the legend. Non-excess stars ($<5\sigma'$; 2,185 stars) are plotted as grey points. The dashed lines represent our computed polynomials used to trace the main sequence M dwarfs. \label{fig:colorcolorcriteria}}
\end{figure}

	One complication with using absolute magnitude as a temperature proxy is that pre-main sequence stars will appear as earlier spectral types (than their main sequence counterparts at the same mass) due to their larger photospheres. A12 discovered that $V - K_s$ was a much better discriminator for $T_\mathrm{eff}$. The SDSS, $r-z$ color has been shown to be a reliable proxy for stellar effective temperature \citep[e.g.,][]{bochanski:2010:2679,bochanski:2011:98}. To overcome the limitations of using the A12 absolute magnitude polynomials, we examined $r - z$ color as a function of \textit{WISE} colors for the 309 remaining stars in our sample, and the stars not exhibiting \textit{WISE} excesses ($ < 5\sigma'$). We computed $r-z$ polynomials, similar to the methods A12 used for $V-K_s$ color. In $r-z$ space, two distinct populations become apparent at $W1-W3 \approx 1.5$ and $W2-W3 \approx 1$, as is shown in Figure~\ref{fig:colorcolorcriteria}. Using these criteria as \textit{WISE} color cuts, we kept only stars that fell below the two previously stated colors for computing $r-z$ polynomials, giving us 2,209 and 2,042 stars for each cut, respectively. A12 used a sigma-clipping method to derive their fourth order polynomials of the form,
$$
color(p) = a_0 + a_1 p + a_2 p^2 + a_3 p^3 + a_4 p^4,
$$	
where $p$ is the effective temperature proxy used, in our case $r-z$ color. We chose to implement a Bayesian framework to compute our fourth order polynomial. To estimate each coefficient for our polynomial we used \texttt{emcee}\footnote{http://dan.iel.fm/emcee} \citep{foreman-mackey:2013:306}, a Python implementation of the affine-invariant Markov Chain Monte Carlo (MCMC) ensemble sampler \citep{goodman:2010:65}. Our coefficients, taken as the 5$^\mathrm{th}$ percentile values from \texttt{emcee}, are listed in Table~\ref{tbl:polyvals} and the polynomials are shown in SDSS and \textit{WISE} color-color space in Figure~\ref{fig:colorcolorcriteria}. Using the same definition for $\sigma'$, we recomputed values for our sample and kept only stars with $\sigma' \ge 5$. After these cuts we were left with \finalcut\ stars showing excess flux at 12 \um. 
	
\begin{deluxetable*}{ccccccc}
\tabletypesize{\scriptsize}
\tablecolumns{7}
\tablewidth{0pt}
\tablecaption{$r-z$ Polynomial Coefficents\label{tbl:polyvals}}
\tablehead{
\colhead{Color} & \colhead{$a_0$} & \colhead{$a_1$} & \colhead{$a_2$} & \colhead{$a_3$} & \colhead{$a_4$} & \colhead{$\chi^2$} 
}
\startdata
$W1-W3$ & $0.0872^{+0.2238}_{-0.2127}$ & $0.5527^{+0.3883}_{-0.4223}$ & $-0.2869^{+0.2259}_{-0.1635}$ & $0.06287^{+0.06699}_{-0.07006}$ & $-0.004410^{+0.006476}_{-0.006269}$ & 1.3118 \\
$W2-W3$ & $0.3776^{+0.2069}_{-0.1972}$ & $-0.04019^{+0.37338}_{-0.38995}$ & $-0.0054^{+0.2829}_{-0.2112}$ & $0.002268^{+0.065797}_{-0.068222}$ & $0.000299^{+0.006427}_{-0.006227}$ & 1.0543 
\enddata
\end{deluxetable*}	

	Any color selection criterion omitting $W4$ has the possibility of excluding stars with cold dust populations that peak at wavelengths longer than 12 \um. Therefore, we examined all stars with a SNR$_{W4} \ge 3$ prior to exclusion. We did not have enough stars with high S/N $W4$ measurements to recompute polynomials using $r-z$ color for excess $W4$ detections. Instead, we used the A12 polynomials for $W1-W4$ and $W3-W4$ and found that all \wfoursample\ stars with a SNR$_{W4} \ge 3$ exhibited levels of MIR flux far greater than estimated photospheric levels ($\gg 5\sigma'$). Every star that showed an excess at 22 \um\ also showed an excess at 12 \um. 
	
	As a final quality check we inspected each candidate within the \textit{WISE} image archives. For each candidate, we assigned a quality flag, defined in Table~\ref{tbl:quality}, with \textsc{quality} $= 1$ representing the highest quality candidates and \textsc{quality} $= 4$ representing the lowest quality candidates. These quality flags are available in the online catalog under the heading `\textsc{quality}'. We will refer to this sample of \totalsample\ disk candidates as the ``combined sample" throughout this study.
	
\begin{deluxetable*}{clc}
\tabletypesize{\scriptsize}
\tablecolumns{3}
\tablewidth{0pt}
\tablecaption{\textit{WISE} Quality Flags\label{tbl:quality}}
\tablehead{
\colhead{Flag} & \colhead{Description} & \colhead{Number}
}
\startdata
1 & Clear 12 and 22 \um\ source. & 5\\
2 & Clear 12 \um\ source with a 22 \um\ source that may be: affected by contamination, slightly offset, or low S/N. & 11\\
3 & Clear 12 \um\ source with no obvious 22 \um\ source. & 54\\
4 & 12 \um\ source that may be: affected by contamination, slightly offset, or low S/N. & 105
\enddata
\end{deluxetable*}
	
\begin{figure}
 \plotone{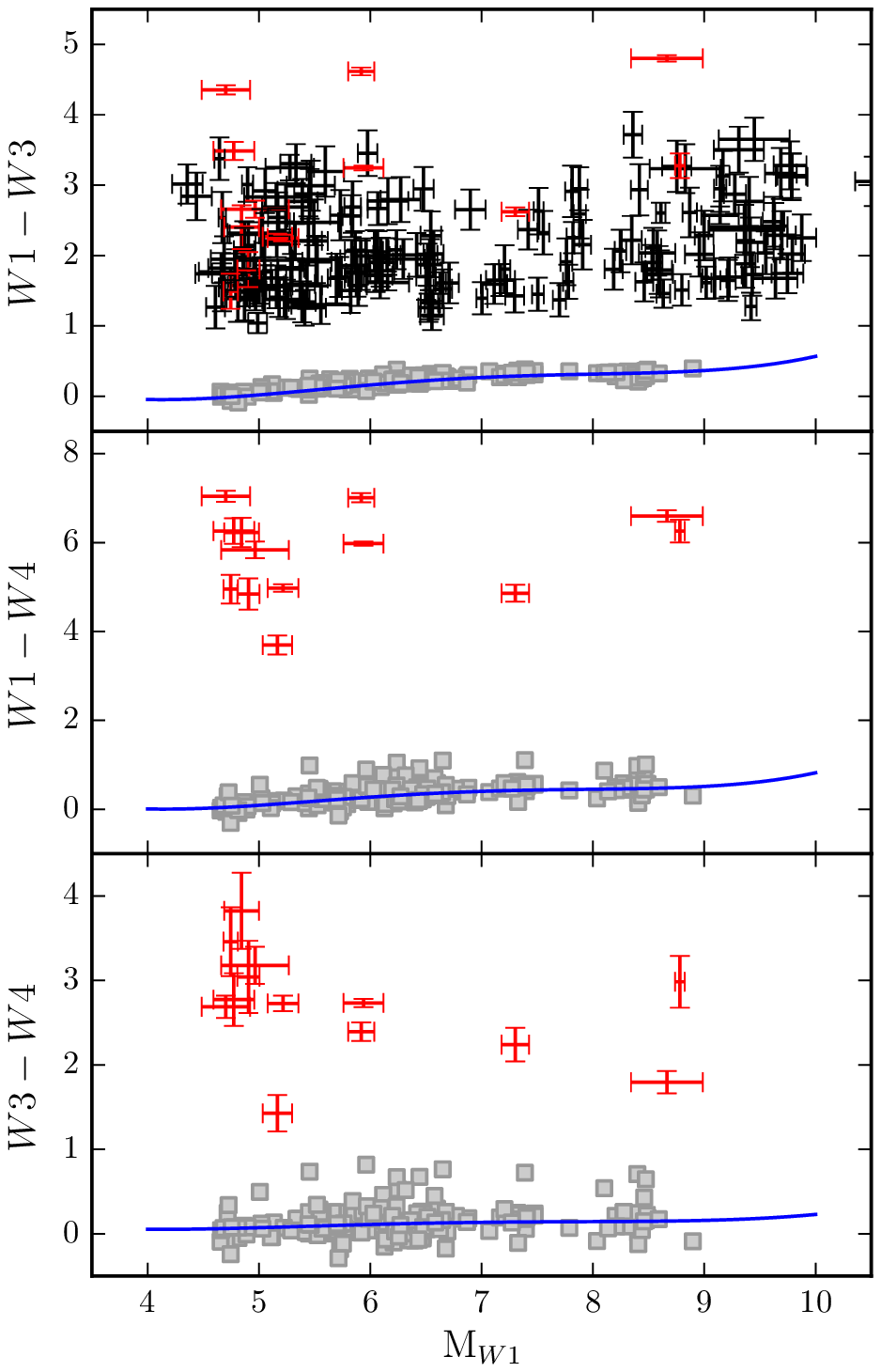}
\caption{\textit{WISE} color-magnitude diagrams. Disk candidates from this study (pluses) 
and the nearby K and M dwarfs \citep[squares;][]{reid:1997:2246,reid:2004:463,reid:2007:2825} are shown. 
Only stars with SNR$_{W4} \geq 3$ and \textsc{cc\_flags}$_{W4}=0$ were used in the $W1-W4$ and $W3-W4$ plots (red pluses). 
Overplotted are the main sequence color polynomials from A12 (solid lines). Both populations are well separated indicating that
the vast majority of these stars likely harbor dust populations. Even if distance errors were as large as 50\%, which would move the points to the 
left/right on each plot, the excesses we observe would still be statistically significant.\label{fig:avenhaus12}}
\end{figure}

	Figure~\ref{fig:avenhaus12} shows the \textit{WISE} color-magnitude diagram for our candidates stars showing IR excesses, and also nearby K- and M-type dwarfs \citep[][]{reid:1997:2246,reid:2004:463,reid:2007:2825}, a statistically older population of stars without known disks. The A12 polynomial describing the main sequence is over plotted (blue solid line). The population of nearby stars distinctly follows the A12 polynomials.

	If we assume that for a given $r-z$ color the spread of intrinsic $W1-W3$ color for the non-IR excess stellar population follows a symmetric distribution, then it is possible that some of our candidates are false-positives within the red tail of the distribution. Assuming that \textit{WISE} should have been able to achieve $5\sigma$ sensitivity in $W3$ for any star with $J\leq17$ ($\sim$0.177 mJy) in the W11 catalog (57,339 stars), then we expect less than one false positive at $\sigma' = 5$. Most of the stars in both our samples are well above the $5\sigma'$ level, therefore we expect all of our candidates to exhibit real excesses not attributable to intrinsic photospheric levels. We list the excess significances ($\sigma'$) in Table~\ref{tbl:params} and in the online catalog.

\begin{figure}
 \plotone{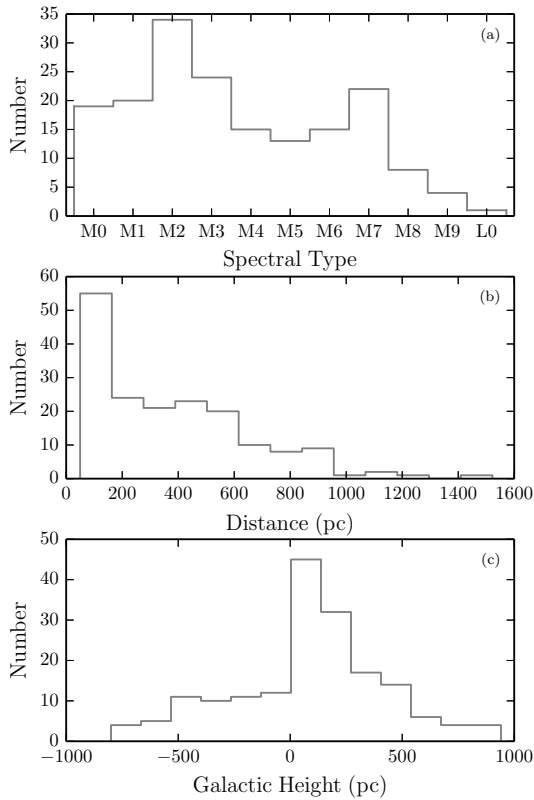}
\caption{Distributions of the combined sample: (a) spectral type distribution; (b) distance distribution; 
(c) distribution of vertical distance away from the Galactic plane. The majority of stars are earlier 
spectral types ($< \textrm{dM5}$) within $d<600$ pc and $|Z|<500$ pc. \label{fig:dist}}
\end{figure}

\begin{figure*}
 \plotone{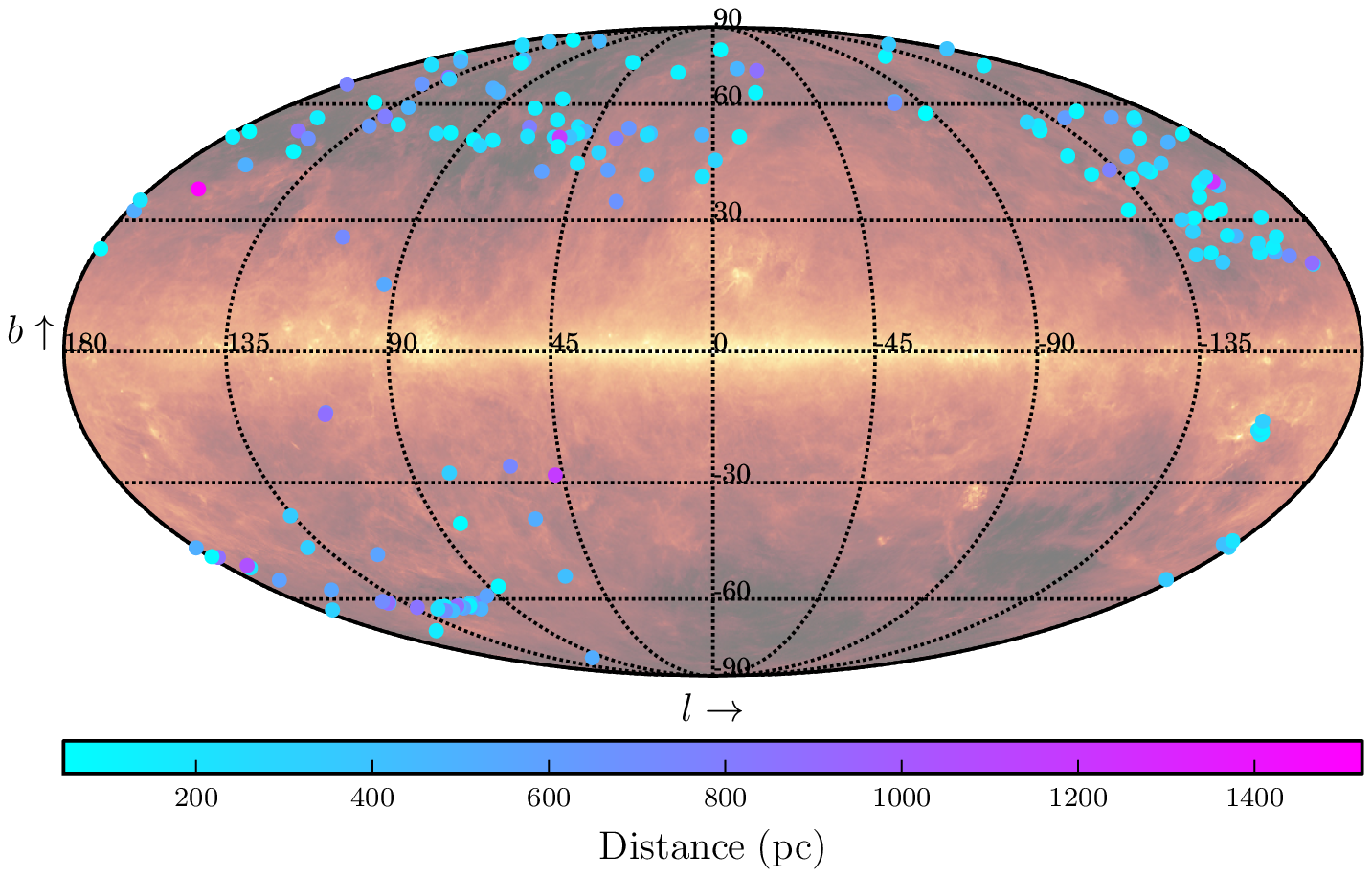}
\caption{Galactic coordinates and distances of disk candidates from this study. Stars are color-coded by distance.
The points are plotted on the \textit{IRAS}/\textit{COBE} 100 \um\ dust map \citep{schlegel:1998:525}. 94\% of the sample is located off the Galactic plane ($|b| > 20^\circ$). \orionsample\ of the 11 stars with $|b| \le 20^\circ$ are found within nearby proximity of the star-forming Orion OB1 association ($l \approx -156^\circ$, $b \approx -16^\circ$, $d \approx 200$--500 pc; \citealt{bally:2008:459}).
\label{fig:galactic}}
\end{figure*}

	Coordinates, distances, and spectral types are listed in Table~\ref{tbl:params}. One of our stars was re-spectral typed from an M9 to an L0 in \citet{schmidt:2010:1808}, however, we chose not to remove this star from our sample. The distributions of spectral types, distances, and height above the Galactic plane are shown in Figure~\ref{fig:dist}. Galactic heights were estimated following \citet{bochanski:2010:2679} with $R_\odot=8.5$ kpc \citep{kerr:1986:1023} and $Z_\odot=15$ pc above the Galactic plane \citep{cohen:1995:874,ng:1997:65,binney:1997:365}. The majority of stars in our samples are found within 500 pc from the Galactic plane. In Figure~\ref{fig:galactic} we show the locations and distances of the sample plotted over a dust map, created using data from the \textit{Cosmic Microwave Background Explorer} \citep[\textit{COBE;}][and references therein]{boggess:1992:420} and \textit{IRAS} \citep{schlegel:1998:525}. We have \orionsample\ candidate stars within the footprint of the Orion OB1 association \citep{warren:1977:115}. Although there appears to be structure in Figure~\ref{fig:galactic}, this is due to the SDSS sky coverage which is observed in wide 2.5$^\circ$ stripes, and all our candidates are spread throughout the SDSS DR7 footprint\footnote{http://www.sdss.org/dr7/coverage/}.
	
	To test for additional spatial structure among our candidates, we wrote a simple friends-of-friends algorithm \citep[FoF;][]{huchra:1982:423}. We used the most general method for finding groups and used a distance discriminator to determine if two stars were ``friends," arbitrarily choosing a critical length scale of 10 pc. The size is somewhat arbitrary depending on whether members make up: 1) a cluster, in which case members will be closely grouped in position; or 2) a moving group, in which case members will be closely grouped in velocity space. We only tested for spatial structure corresponding to the distribution of candidates throughout the Galaxy. Using 3-dimensional position vectors, made by combining photometric parallax distances and the positions of our stars, we were able to group stars together by their positions within the Galaxy. With our method we found two stellar groupings (available in the online catalog), one of which was centered on Orion. Neither of these stellar groupings appeared to have significantly similar kinematics (kinematics computed in \S\ref{kinematics}), and are likely not members of a moving group.

\begin{deluxetable*}{lrrcccccccc}
\tabletypesize{\scriptsize}
\tablecolumns{11}
\tablewidth{0pt}
\tablecaption{Disk Candidate Parameters\label{tbl:params}}
\tablehead{
& \colhead{R.A.} & \colhead{Decl.} & & & & & & & \colhead{} &\\
\colhead{Catalog} & \colhead{(J2000.0)} & \colhead{(J2000.0)} & \colhead{SpT\tablenotemark{a}} & \colhead{$d$\tablenotemark{b}} & \colhead{Z} & \colhead{$A_V$\tablenotemark{c}} & \colhead{$R_V$\tablenotemark{d}} & \colhead{$\sigma'_\mathrm{W1-W3}$\tablenotemark{e}} & \colhead{$\sigma'_\mathrm{W1-W4}$\tablenotemark{e}} & \colhead{$\sigma'_\mathrm{W3-W4}$\tablenotemark{e}} \\
\colhead{Number} & \colhead{(deg)} & \colhead{(deg)} & \colhead{$\pm 1$} & \colhead{(pc)} & \colhead{(pc)} &\colhead{(mags)} & & \colhead{} &&
}
\startdata
40.......... & 83.032543 & 0.284724 & M2 & 195 & -44 & 0.52 $\pm$ 0.02 & 3.1 & 63.7 & 21.4 & 8.8 \\
42.......... & 83.296678 & 0.232939 & M3 & 128 & -23 & 0.35 $\pm$ 0.06 & 3.1 & 111.6 & 153.4 & 76.3 \\
62.......... & 129.436350 & 21.646841 & M7 & 108 & 73 & 0.00 $\pm$ 0.00 & 3.1 & 119.0 & 60.5 & 17.5 \\
78.......... & 142.779528 & 10.102017 & M6 & 135 & 101 & 0.77 $\pm$ 0.06 & 3.1 & 20.0 & 29.0 & 13.0 \\
108........ & 180.890419 & 49.244348 & M1 & 648 & 607 & 0.00 $\pm$ 0.03 & 3.1 & 8.7 & 17.5 & 9.8 \\
110........ & 181.192639 & 40.432928 & M4 & 475 & 471 & 0.16 $\pm$ 0.05 & 3.1 & 97.3 & 83.7 & 29.0 \\
115........ & 185.176759 & 48.151921 & M0 & 894 & 844 & 0.15 $\pm$ 0.03 & 3.1 & 9.3 & 24.0 & 11.7 \\
126........ & 203.510751 & 10.271275 & M2 & 872 & 836 & 0.13 $\pm$ 0.01 & 3.1 & 81.4 & 73.6 & 27.9 \\
140........ & 232.129207 & 38.051640 & M5 & 150 & 138 & 0.15 $\pm$ 0.11 & 3.1 & 45.9 & 29.8 & 14.9 \\
142........ & 232.736020 & 20.902880 & M1 & 657 & 540 & 0.00 $\pm$ 0.00 & 3.1 & 26.0 & 39.5 & 19.9 \\
152........ & 236.977925 & 52.815983 & M1 & 283 & 227 & 0.00 $\pm$ 0.03 & 3.1 & 56.2 & 73.2 & 40.6 \\
156........ & 238.196155 & 35.302076 & M3 & 1174 & 923 & 0.35 $\pm$ 0.04 & 3.1 & 32.7 & 27.6 & 12.2 \\
164........ & 284.933373 & 78.072385 & M0 & 723 & 332 & 0.00 $\pm$ 0.00 & 3.1 & 7.2 & 19.4 & 11.7 
\enddata
\tablecomments{Values listed only for stars with SNR$_{W4} \geq 3$. Values for the entire sample from this study are available online.}
\tablenotetext{a}{Determined in \citet{west:2011:97}.}
\tablenotetext{b}{Photometric distances have typical uncertainties of $\sim$20\% \citep{bochanski:2010:2679,jones:2011:44}.}
\tablenotetext{c}{Measured in \citet{jones:2011:44}.}
\tablenotetext{d}{Adopted parameters from \citet{jones:2011:44} measurements.}
\tablenotetext{e}{$\sigma'$ values from A12. These represent the IR excess significance in two {\em WISE} bands.}
\end{deluxetable*}
	
\subsection{Interstellar Reddening} \label{reddening}

	The interstellar medium (ISM) contains varying amounts of dust, which affect observations depending on their line-of-sight (LOS) by absorbing or scattering background light. These effects ``redden" or ``extinct" observed fluxes, and are most significant at short wavelengths, falling off in the MIR \citep[e.g.,][]{fitzpatrick:1999:63}. For stars within $\sim$50 pc, extinction effects can typically be ignored due to the lack of dust within the local bubble \citep{lallement:2003:447}. Due to the distances to the stars in this study, effects from interstellar extinction may be significant and should be accounted for.
	
	To correct for the effects of dust, \citet{schlegel:1998:525} created dust maps by measuring galactic extinction using \textit{COBE} and \textit{IRAS}. Although SDSS provides extinction estimates for photometric sources using these maps, caution must be taken due to the fact that the \citet{schlegel:1998:525} dust maps estimate the \textit{total} extinction along a LOS out of the Galaxy. SDSS stars at distances of hundreds of parsecs might have their fluxes reduced by only a fraction of the total extinction measured in these maps, requiring us to estimate extinction not only along a LOS, but as a function of distance. Extinction effects due to circumstellar dust are also possible if these are pre-main sequence stars. However, extinction due to circumstellar dust will only have a significant effect if the star is a protostar ($<1$ Myr) and still harbors an infalling envelope, or the disk is seen edge-on \citep[for a more detailed explanation see][]{hartmann:1998:}. Measured extinction effects cannot be reliably attributed to interstellar material or a circumstellar disk, since the effect is cumulative for all sources of extinction. For the purposes of this paper, we assume the disk material to be optically thin at all observed wavelengths since none of our stars are protostars.
	
	\citet[][hereafter J11]{jones:2011:44} measured the total interstellar extinction, $A_V$, and the ratio of total extinction to reddening, $R_V$, for the W11 catalog. This was done by comparing the spectra for each star in the W11 catalog to high S/N SDSS spectra for nearby M dwarfs at low or zero extinction lines-of-sight. Using the $A_V$ and $R_V$ values from J11, along with relative extinction $A_\lambda / A_V$, the ratio of extinction in a given bandpass to extinction in the $V$-band, measurements from the Asiago Database \citep{moro:2000:361, fiorucci:2003:781}, we calculated the extinction for each SDSS, 2MASS, and \textit{WISE} photometric band. $V$-band extinction is commonly used for relative extinction values in the optical due to extinction laws being expressed in terms of $A_V$ and $R_V$. The Asiago Database contains measured relative extinction values for the SDSS and 2MASS photometric systems (e.g., $A_r / A_V$, $A_{K_s} / A_V$, etc.) using the \citet{fitzpatrick:1999:63} extinction law. To estimate relative extinction values for the SDSS and 2MASS photometric systems, bandpass response curves were convolved over a model M-spectral type photosphere to determine the relative extinction for each bandpass \citep[for more details see][]{moro:2000:361, fiorucci:2003:781}. These relative extinction measurements for the SDSS and 2MASS bandpasses allowed us to estimate specific bandpass extinctions using the $A_V$ values from J11 (e.g., $A_r = ( A_r / A_V ) A_{V_{J11}}$).
	
	For $W1$ and $W2$, we used relative extinction $A_\lambda / A_K$ measurements for \textit{Spitzer}'s Infrared Array Camera \citep[IRAC;][]{fazio:2004:10} channel 1 (3.6 \um) and channel 2 (4.5 \um) from \cite{indebetouw:2005:931}, which are analogous to $W1$ and $W2$ (D. P. Clemens, personal communication). For wavelengths longer than 5 \um\ the extinction curve begins to flatten, and we assumed equal extinction values for $W2$, $W3$, and $W4$. The extinction parameters used for this study are listed in Table~\ref{tbl:extinction}. Extinction corrected photometry is included in the online catalog. The majority of our sample ($\sim$90\%) had relatively small amounts of extinction ($A_V \leq 0.5$), with $V$-band extinction values as high as 3.1 mags. Our typical extinction values are smaller than those measured around nearby star-forming regions \citep[$A_V = 1$--12;][]{luhman:2003:1093,luhman:2004:816,mcclure:2009:l81}, as expected for field stars.

\begin{deluxetable*}{ccccccccccc}
\tabletypesize{\scriptsize}
\tablecolumns{11}
\tablewidth{0pt}
\tablecaption{Adopted Reddening Parameters\label{tbl:extinction}}
\tablehead{
\colhead{$R_V$\tablenotemark{a}} & \colhead{$A_u/A_V$\tablenotemark{b}} & \colhead{$A_g/A_V$\tablenotemark{b}} & \colhead{$A_r/A_V$\tablenotemark{b}} & 
\colhead{$A_i/A_V$\tablenotemark{b}} & \colhead{$A_z/A_V$\tablenotemark{b}} & \colhead{$A_J/A_V$\tablenotemark{b}} & \colhead{$A_H/A_V$\tablenotemark{b}} & 
\colhead{$A_{K}/A_V$\tablenotemark{b}} & \colhead{$A_{W1}/A_{K}$\tablenotemark{c}} & \colhead{$A_{W2}/A_{K}$\tablenotemark{c}}
}
\startdata
2.1 & 2.06 & 1.28 & 0.78 & 0.56 & 0.43 & 0.27 & 0.18 & 0.12 & 0.56 & 0.43\\
3.1 & 1.67 & 1.19 & 0.84 & 0.62 & 0.46 & 0.27 & 0.17 & 0.12 & 0.56 & 0.43\\
5.0 & 1.38 & 1.12 & 0.88 & 0.67 & 0.49 & 0.26 & 0.16 & 0.11 & 0.56 & 0.43
\enddata
\tablenotetext{a}{Adopted $R_V$ was based on measurements from J11: $R_V=2.1$ for $0<R_{V_{J11}}<2.5$; $R_V=3.1$ for $2.5\le R_{V_{J11}}<4$; $R_V=5$ for $4\le R_{V_{J11}}<10$.}
\tablenotetext{b}{Adopted from \citet{moro:2000:361, fiorucci:2003:781} (see text).}
\tablenotetext{c}{Adopted from \citet{indebetouw:2005:931} (see text).}
\end{deluxetable*}

\subsection{Flux Conversions from Magnitudes} \label{flux conversion}
	
	Flux densities for SDSS asinh AB magnitudes were computed following the methods outlined in the DR7 flux calibration primer\footnote{http://www.sdss.org/dr7/algorithms/fluxcal.html}. 2MASS Vega magnitudes were converted to flux densities following the steps outlined in the explanatory documentation\footnote{http://www.ipac.caltech.edu/2mass/releases/second/doc/sec6\_4a.html}. \textit{WISE} Vega magnitudes were converted to flux densities following the all-sky explanatory supplement\footnote{http://wise2.ipac.caltech.edu/docs/release/allsky/expsup/sec4\_4h.html}, assuming a spectral slope of $-2$ ($F_\nu \sim \nu^{-2}$). For extremely red sources (e.g., ultra-luminous infrared galaxies), it was found that $W4$ overestimated fluxes by $\sim$10\% since the stellar sources used to photometrically calibrate \textit{WISE} were significantly bluer in comparison (e.g., A-K dwarfs and K/M giants). Since we expect warm dust to mimic an extremely red source, we applied flux corrections to $W4$, reducing measured fluxes by 10\%. It is also expected that red sources should have their $W3$ fluxes underestimated by $\sim$10\%, however, we chose not to apply this correction since the shape of the SED at $W3$ is not known a priori.

\section{Evaluating the Disk Candidates}\label{methods}

	To properly characterize the dust content for each star, we first estimated the expected flux of the star using model stellar photospheres. We used a pre-computed grid of BT-Settl models \citep{allard:2012:3, allard:2012:2765}, computed with the PHEONIX atmosphere code \citep{hauschildt:1999:377, allard:2001:357}, to determine stellar photospheric flux with which to fit optical and NIR photometry. For each star, we used grids spanning $T_\mathrm{eff}$ between $\pm$1,000 K of nominal values, determined by using spectral types from W11 and $T_\mathrm{eff}$ values from \citet[]{reid:2005:}, in increments of 100 K. The other parameters spanned by the grids were log($g$) in the range [0.5, 6.0] in increments of 0.5 dex, and metallicity [M/H] in the range [$-2.5$, 0.5] in increments of 0.5 dex. All of the observed photometry was color-corrected for extinction effects (\S\ref{reddening}). 

	To fit the stellar photosphere models to our photometry, we computed a $\chi^2$ minimization as a function of the effective temperature, surface gravity, and metallicity. We performed two separate fits, one to the photometry and one to the SDSS spectrum. We gave equal weights to both the photometric and spectral fits because we found that separate fits typically agreed to within $\pm$100 K. The SDSS spectra were radial velocity and extinction corrected using J11 extinction values (\S\ref{reddening}) and the extinction law from \citet{fitzpatrick:1999:63} prior to fitting. The spectrum and model photosphere were normalized to their values at 7500 \AA. The model photosphere was then resampled to SDSS wavelengths while conserving flux using the Python package {\it Pysynphot}\footnote{http://stsdas.stsci.edu/pysynphot/}. 
	
\begin{figure}
 \plotone{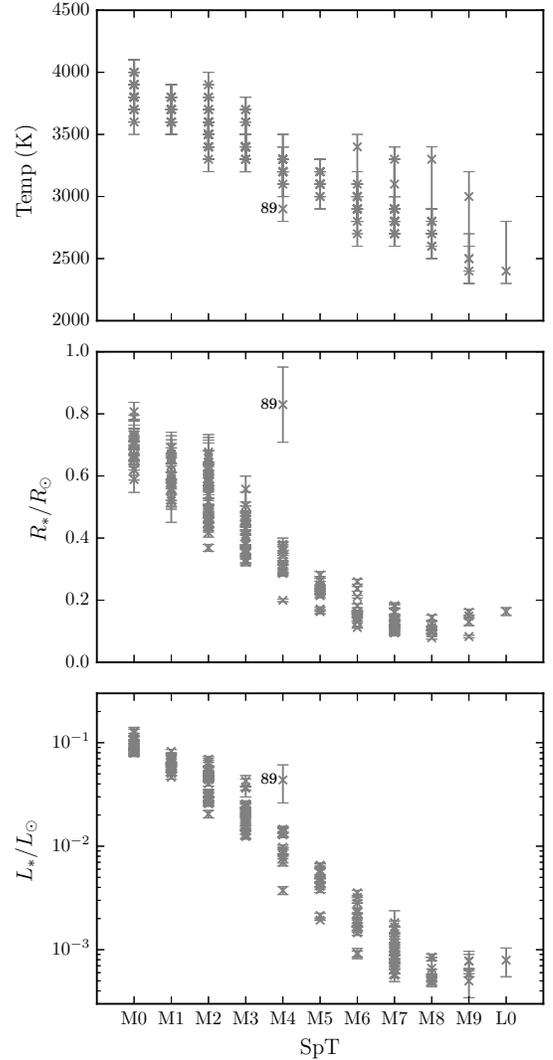}
\caption{Best fit parameters from our SED fitting method as a function of spectral type for our combined sample. 
The majority of stars appear to follow the expected trends for temperature, luminosity, and radius. 
The outlier in the M4 spectral type (marked by its candidate number) will be 
discussed in \S\ref{ttauri}.
\label{fig:modelparams}}
\end{figure}

	For the photometric fit, the modeled photospheric flux was integrated over the \textit{ugriz} and \textit{JHK}$_s$ relative spectral response curves (RSRs) to produce synthetic magnitudes, which we matched to our observed photometry for the $\chi^2$ minimization. \textit{WISE} bands were omitted due to expected deviations from the stellar photosphere in $W3$ and $W4$ and possible deviations in $W1$ and $W2$ for pre-transitional disk structures \citep{espaillat:2007:l135,espaillat:2011:49}. The 2MASS RSRs were taken from \citet{cohen:2003:1090}, \textit{WISE} RSRs from \citet{wright:2010:1868}, and SDSS RSRs from \citet{doi:2010:1628}. All RSRs were normalized to unity at the peak value of each band. Each photometric band was weighted by the inverse variance of the photometric values. Synthetic measurements were normalized to the observed \textit{K$_s$} magnitude for fitting. The results of our fitting method can be seen in Figure~\ref{fig:modelparams}. The average errors in our fits were $\sigma_{T_\mathrm{eff}}<200$ K, $\sigma_\mathrm{[M/H]}<1$, and $\sigma_{\mathrm{log(}g)}<1$. 
	
	The primary purpose of our stellar model fits was to estimate the stellar SEDs and ascertain the amount of excess IR flux, and not to derive robust fundamental stellar parameters. The shape of the SED is dependent primarily on the stellar effective temperature ($T_\mathrm{eff}$), and less so on the metallicity or surface gravity. Allowing all three parameters to float gave us better estimates on $T_\mathrm{eff}$ and the uncertainty in our fit. The best-fit $T_\mathrm{eff}$ for our candidates are listed in Table~\ref{tbl:modelparams}, and we include all model parameters and chi-squared values in the online catalog. There is one clear outlier in the M4 spectral type bin, possibly due to enlarged photospheres owing to a young age. This star will be discussed further in \S\ref{ttauri}. 
	
	For completeness, we also include the traditional $\chi$ metric for measuring excesses in the IR defined as,
\begin{equation}
\chi_{\lambda} = \frac{F_\lambda - F_{\ast,\lambda}}{\sigma_\lambda},
\end{equation}
where $F_\lambda$ is the observed flux at wavelength $\lambda$, $F_{\ast,\lambda}$ is the expected stellar flux at wavelength $\lambda$, and $\sigma_\lambda$ is the uncertainty in the measurement at wavelength $\lambda$. This method depends only on estimating $T_\mathrm{eff}$, which our above process was able to do within a few hundred kelvin for most of our stars (e.g., see Figure~\ref{fig:SED}). Within our combined sample, all stars were found to have $\chi_{12} > 2$, and every star with a SNR$_{W4} \geq 3 $ had $\chi_{22} > 3$. We chose not to remove any candidates with $\chi_{12} \leq 3$ from our sample as these stars all had higher significances using the method of A12 (\S\ref{identifying}), which is a purely empirical method that does not depend on stellar models. These values are included in the online catalog, along with flux densities for \textit{WISE} observations and expected stellar photospheric levels.

\begin{deluxetable*}{lcccccccc}
\tabletypesize{\scriptsize}
\tablecolumns{9}
\tablewidth{0pt}
\tablecaption{Model Parameters\label{tbl:modelparams}}
\tablehead{
\colhead{Catalog} & \colhead{$T_\ast$\tablenotemark{a}} & \colhead{$L_\ast$} & \colhead{$R_\ast$} & \colhead{$T_\mathrm{dust}$} & \colhead{$L_\mathrm{dust}/L_\ast$\tablenotemark{b} } & \colhead{$D_\mathrm{min}$\tablenotemark{b} } & \colhead{$M_\mathrm{dust}$\tablenotemark{b}} & \colhead{$\chi^2_\mathrm{fits}$\tablenotemark{c}}\\
\colhead{Number} & \colhead{(K)} & \colhead{($10^{-2}L_\odot$)} & \colhead{($R_\odot$)} & (K) & & (AU) & ($M_\mathrm{Moon}$) &
}
\startdata
40.......... & $3600^{+100}_{-100}$ & 6.24 $\pm$ 0.52 & 0.64 $\pm$ 0.04 & $325^{+15}_{-15}$ & $3.5\times10^{-2}$ & 0.18 & $2.0\times10^{-5}$ & 33.85 \\[4pt]
42.......... & $3400^{+200}_{-100}$ & 1.75 $\pm$ 0.15 & 0.38 $\pm$ 0.03 & $185^{+10}_{-10}$ & $> 2.7\times10^{-1}$ & 0.30 & $4.0\times10^{-4}$ & 321.87 \\[4pt]
62.......... & $2800^{+600}_{-100}$ & 0.10 $\pm$ 0.03 & 0.14 $\pm$ 0.02 & $290^{+10}_{-5}$ & $> 5.9\times10^{-1}$ & 0.03 & $8.5\times10^{-6}$ & 28.39 \\[4pt]
78.......... & $3000^{+100}_{-100}$ & 0.19 $\pm$ 0.01 & 0.16 $\pm$ 0.00 & $165^{+20}_{-20}$ & $2.3\times10^{-1}$ & 0.12 & $6.0\times10^{-5}$ & 17.53 \\[4pt]
108........ & $3600^{+100}_{-100}$ & 6.58 $\pm$ 0.32 & 0.66 $\pm$ 0.03 & $155^{+25}_{-35}$ & $6.3\times10^{-2}$ & 0.83 & $7.2\times10^{-4}$ & 22.26 \\[4pt]
110........ & $3200^{+200}_{-100}$ & 1.36 $\pm$ 0.13 & 0.38 $\pm$ 0.02 & $220^{+5}_{-10}$ & $\sim 1$ & 0.19 & $5.8\times10^{-4}$ & 15.48 \\[4pt]
115........ & $3900^{+100}_{-100}$ & 9.20 $\pm$ 0.38 & 0.67 $\pm$ 0.05 & $125^{+20}_{-25}$ & $2.0\times10^{-1}$ & 1.51 & $7.5\times10^{-3}$ & 24.99 \\[4pt]
126........ & $3600^{+100}_{-100}$ & 4.03 $\pm$ 0.22 & 0.52 $\pm$ 0.05 & $195^{+5}_{-10}$ & $> 9.8\times10^{-1}$ & 0.41 & $2.7\times10^{-3}$ & 37.82 \\[4pt]
140........ & $3200^{+100}_{-100}$ & 0.64 $\pm$ 0.03 & 0.26 $\pm$ 0.01 & $220^{+10}_{-10}$ & $8.2\times10^{-2}$ & 0.13 & $2.2\times10^{-5}$ & 42.21 \\[4pt]
142........ & $3700^{+100}_{-100}$ & 5.73 $\pm$ 0.51 & 0.58 $\pm$ 0.08 & $155^{+10}_{-10}$ & $1.7\times10^{-1}$ & 0.77 & $1.7\times10^{-3}$ & 23.98 \\[4pt]
152........ & $3600^{+100}_{-100}$ & 5.45 $\pm$ 0.30 & 0.60 $\pm$ 0.04 & $175^{+10}_{-10}$ & $6.4\times10^{-2}$ & 0.59 & $3.7\times10^{-4}$ & 10.50 \\[4pt]
156........ & $3700^{+100}_{-100}$ & 3.63 $\pm$ 0.17 & 0.47 $\pm$ 0.04 & $185^{+15}_{-15}$ & $> 4.9\times10^{-1}$ & 0.43 & $1.5\times10^{-3}$ & 15.56 \\[4pt]
164........ & $3800^{+100}_{-100}$ & 8.37 $\pm$ 0.43 & 0.67 $\pm$ 0.02 & $130^{+20}_{-30}$ & $7.1\times10^{-2}$ & 1.33 & $2.1\times10^{-3}$ & 11.21 
 \enddata
\tablecomments{Values listed only for stars with SNR$_{W4} \geq 3$. Values for the combined sample are available online.}
\tablenotetext{a}{Upper and lower limits derived from the 1$\sigma$ spread in reduced $\chi^2$ values.}
\tablenotetext{b}{Typical uncertainties are on the order of $\sim$20\%}
\tablenotetext{c}{To estimate the uncertainty in our fits we used a $\chi^2 - \chi^2_\mathrm{min}$ procedure similar to \citet{mohanty:2010:1138}.}
\end{deluxetable*}

\subsection{Binarity}\label{binarity}

	We investigated the possibility that the IR excesses in our stars are due to ultracool companions. To estimate the expected observational signature of ultracool companions, we created combined SEDs for low-mass binaries. To properly scale the photospheric flux density for the M dwarf and the ultracool companion, we used the luminosity-$T_\mathrm{eff}$ relations from \citet{reid:2005:} for M dwarfs, and the maximum values from \citet{baraffe:2003:701} for ultracool companions. To scale the model flux density to an observed flux density, we integrated the flux density for each model (star and companion), and scaled the flux density to the expected luminosity from the literature at an arbitrary distance. The distance falls out of the equation due to the fact that we compute ratios of fluxes densities over integrated bandpasses. Mathematically, this is shown as:
\begin{equation}
F_{\lambda \mathrm{, scaled}} = \frac{C L_{\odot}}{4 \pi d^2 \int_0^\infty B_{\lambda}\, \mathrm{d}\lambda}\times F_{\lambda \mathrm{, model}},
\end{equation}
where $C$ is the constant used to scale the luminosity with effective temperature (or spectral type) taken from \citet{reid:2005:} and \citet{burrows:1997:856}, $d$ is an arbitrary distance, $B_{\lambda}$ is the blackbody flux density at the effective temperature of the star or ultracool companion, and $F_{\lambda \mathrm{, model}}$ is the model flux density.

\begin{figure*}
 \plotone{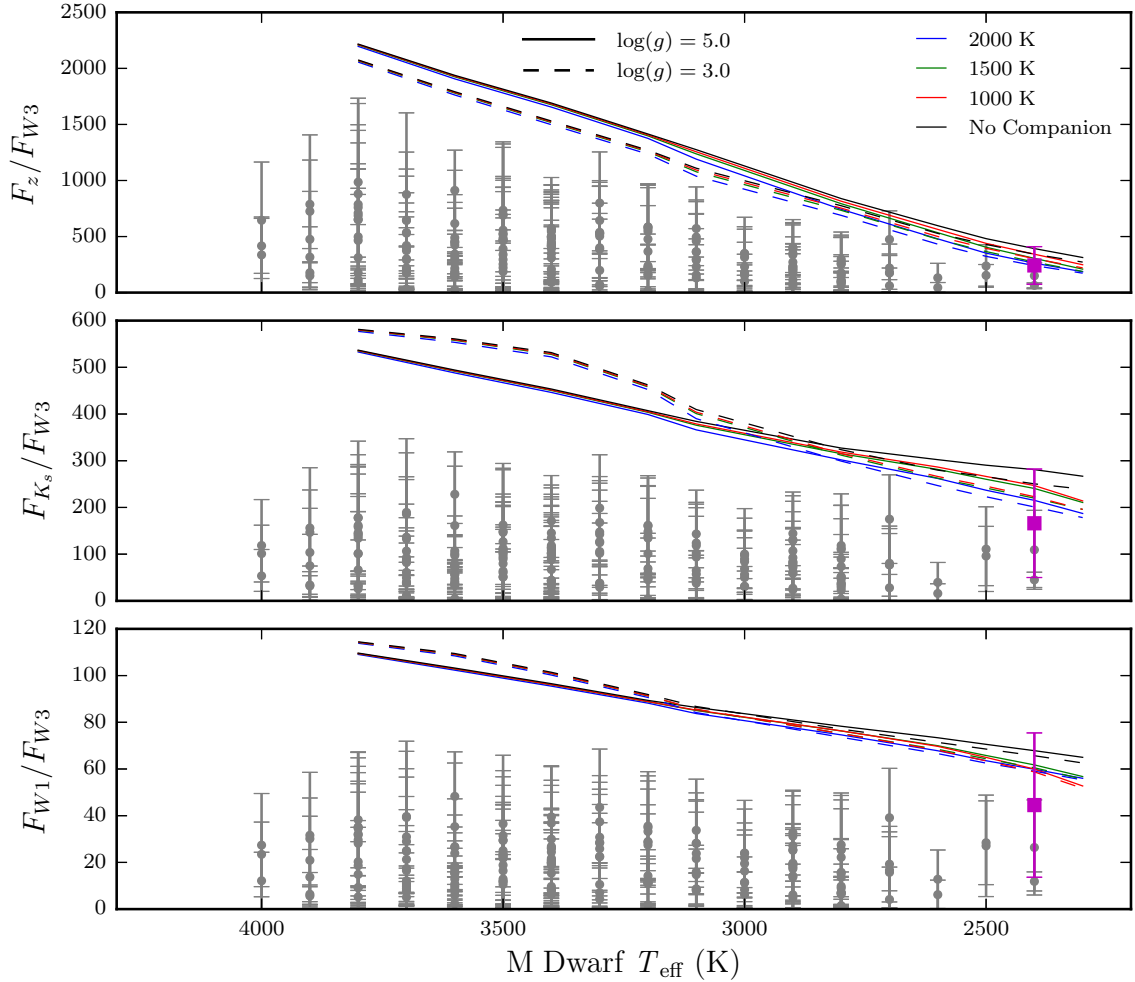}
\caption{Flux density ratios for dM+ultracool companion binary cases. Flux density ratios are taken from SEDs modeled as coadded SEDs of an M dwarf and an ultracool companion. This combined SED represents what is expected from a spectroscopic binary made-up of an M dwarf and an ultracool companion. Each line represents a model of the flux density ratio between two bands for an M dwarf with an ultracool companion (colored lines) and a single M dwarf with no companion (black lines). M dwarf temperatures are listed on the x-axis and companion temperatures are listed in the top right. For the M dwarf, we considered two separate surface gravities, represented by the solid (log($g$) $=5$) and dashed lines (log($g$) $=3$). Stars from our combined sample are plotted (gray circles), with error bars representing the 3$\sigma$ limits. Stars that fall below the model lines have more 12 \um\ flux than the binary case, and thus an IR excesses larger than can be attributed to an ultracool companion. In the W1/W3 flux ratio, all of the stars exceed the IR excess flux attributable to an ultracool companion at $> 3$$\sigma$ confidence. Also plotted is the low-mass binary (M9/T5; magenta square) from \citet[][]{burgasser:2012:110}. Our model is able to reproduce this binary within its 3$\sigma$ uncertainty. \label{fig:binarity}}
\end{figure*}

	We used solar metallicity for both the M dwarf and ultracool companion, log($g$) $=5$ for the ultracool companion, and log($g$) $=3$ \& 5 for the M dwarf to produce a range of values. After scaling each photospheric flux density, we co-added the models and performed synthetic photometry on the combined model to produce flux ratios using the $zK_sW1$-bands relative to the $W3$-band. We then used these synthetic flux density ratios to compare against our observed flux density ratios and determine if any of our excesses could be explained by an ultracool companion. Our model values are shown in Figure~\ref{fig:binarity}. Although two of our candidates fell within the range of harboring an ultracool companion in the $F_z/F_{W3}$ ratio, they did not meet the threshold in other flux ratios, indicating that their excesses are likely not originating from a companion. Focusing on the $F_{W1}/F_{W3}$ ratios, we find that all of our sources have IR excesses much larger than those attributable to an ultracool companion by $\gg 3\sigma$. To test the sensitivity of our model, we applied our method to a known dM+ultracool binary (M9/T5) from \citet[][]{burgasser:2012:110}, and found that our method effectively detected this low-mass binary within the uncertainties (also shown in Figure~\ref{fig:binarity}).

\subsection{Galactic and Extragalactic Contamination}\label{contamination}

	Recent studies of cold disk populations ($T\approx22$ K) using \textit{Herschel} have investigated the possibility that detections of IR excess may be due to IR cirrus/Galactic background or extragalactic contamination \citep[e.g.,][]{eiroa:2011:l4,krivov:2013:32,gaspar:2014:33}. To ensure our excesses are not simply chance alignments of background objects, we must consider and account for both Galactic and extragalactic sources of contamination.
	
\subsubsection{Galactic Background Contamination}
	
	\citet[][hereafter KW12]{kennedy:2012:91} investigated \textit{WISE} excesses for stars in the \textit{Kepler} field-of-view (FOV). Due to the close proximity of the \textit{Kepler} FOV to the Galactic plane, KW12 found a considerable gradient in the number of stars exhibiting $W3$ excesses as a function of distance from the plane, with larger excess source counts closer to the Galactic plane. They attributed this gradient to higher levels of background contamination closer to the Galactic plane, where the ISM has a higher density. Using the Improved Processing of the \textit{IRAS} Survey \citep[IRIS;][]{miville-deschenes:2005:302} 100 \um\ maps\footnote{http://www.cita.utoronto.ca/~mamd/IRIS/}, KW12 determined empirically that stars within a 100 \um\ background level greater than 5 MJy steradian$^{-1}$ were prone to Galactic contamination. A major concern of KW12 was that the \textit{WISE} estimated background level (\textsc{w3sky}) was smooth and did not trace the clumping of the $W3$ excesses seen within the \textit{Kepler} FOV. KW12 used the \textit{WISE} all-sky source catalog, which has been found to {\em underestimate} the background level in $W2$, $W3$, and $W4$\footnote{http://wise2.ipac.caltech.edu/docs/release/allwise/expsup/sec2\_3a.html}, and hence {\em overestimate} the measured flux in these bands for the source\footnote{http://wise2.ipac.caltech.edu/docs/release/allwise/expsup/sec5\_3biii.html}. The AllWISE source catalog used for our study employs a new method to estimate background noise levels, reducing the number of overestimated flux measurements in $W2$, $W3$, and $W4$.
	
\begin{figure}
 \plotone{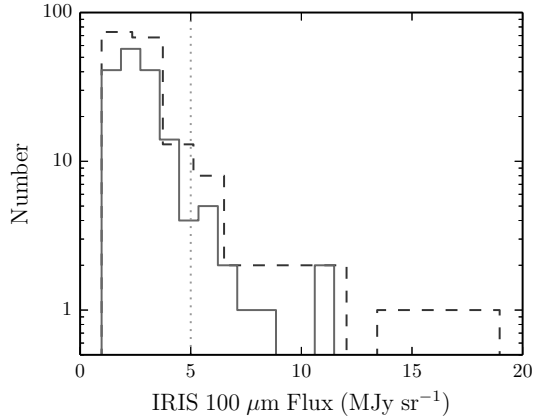}
\caption{Source counts as a function of IRIS 100 \um\ flux level. Shown are our combined sample (dashed line),
and our combined sample without the Orion candidates (solid line). The majority of our sources fall below 
the 5 MJy steradian$^{-1}$ cut used by KW12 (dotted line). The Orion candidates make up a significant
fraction of the stars that fall above 5 MJy steradian$^{-1}$. \label{fig:IRIS}}
\end{figure}

	The majority of W11 stars were sampled at high Galactic latitudes, however, there were a few areas sampled close to the Galactic plane. To test for contamination we used the IRIS 100 \um\ maps to determine the background levels for our stars. We found that the majority of stars fell below the KW12 cut of 5 MJy steradian$^{-1}$, as shown in Figure~\ref{fig:IRIS}. Most of the candidates with a background level higher than 5 MJy steradian$^{-1}$ were found in Orion, which we expect to have a high background. Overall, we do not find an overabundance of sources within areas of high background flux, as seen by KW12. This indicates that the AllWISE source catalog has a robust method for determining background noise levels, significantly reducing the number of false positive excess sources within the catalog. We conclude that our observed excesses do not originate from Galactic background contamination, and did not to remove any candidates falling within regions of high 100 \um\ IRIS fluxes.
	
\subsubsection{Extragalactic Contamination}\label{extragalactic}

	Due to the large FWHM of the \textit{WISE} 12 \um\ (6.5\arcsec) and 22 \um\ (12\arcsec) PSFs, contamination through chance alignment with an extragalactic source is a serious concern and must be addressed. A chance alignment with an object that is bright in the IR, but obscured in the optical (e.g., ultra-luminous infrared galaxies) could contribute large amounts of IR flux. We used two different methods to test the likelihood of such a chance alignment.

	The \textit{WISE} photometric pipeline departs from classical photometric methods by performing profile-fit photometry simultaneously across all bands\footnote{http://wise2.ipac.caltech.edu/docs/release/allsky/expsup/sec4\_4c.html}. This method is expected to detect fainter sources, and reduce confusion. However, without correlating extracted source positions between each band, there is a contamination risk from a background source that was faint at 3.4 and 4.6 \um, but relatively bright at 12 and/or 22 \um.

\begin{figure}
 \plotone{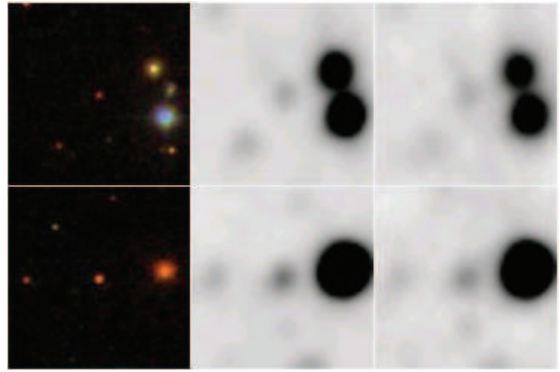}
\caption{1\arcmin\ $\times$ 1\arcmin\ images in SDSS $gri$ (left), $W1$ (middle), and $W2$ (right) for 
objects that failed SExtractor extraction at 4.6 \um. Nearby bright sources coupled with faint detections
 likely confuse SExtractor.\label{fig:sextractor1}}
\end{figure}

\begin{figure}
 \plotone{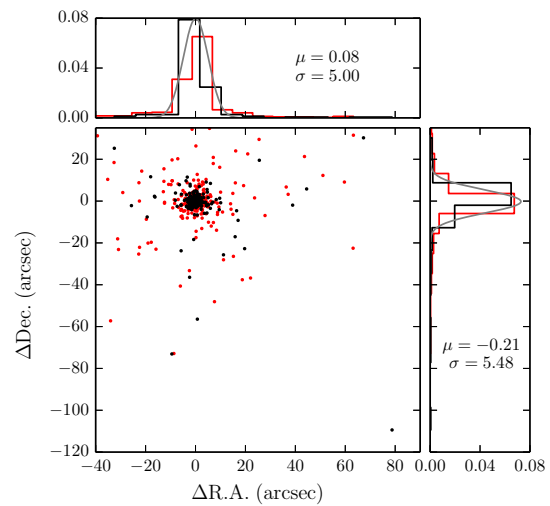}
\caption{Difference in positions between sources extracted from the 3.4 and 12 \um\ bands for our sample (black) and the subsample of SDSS DR10 quasars (red). The best-fit normal distribution to our sample is plotted in gray (distribution parameters listed). For both distributions (stars and QSOs), there is a discrepancy between the 3.4 \um\ and 12 \um\ source positions for many of the sources extracted by SExtractor (see text). 
\label{fig:wisedist}}
\end{figure}

	We chose to employ a more traditional source detection technique by acquiring \textit{WISE} archive images and performing source extraction via SExtractor \citep{bertin:1996:393}. For sources that are detectable in $W1$, $W2$, and $W3$ bands, we expect small offsets between their extracted positions within each separate band since they have similar PSF FWHMs (6.1\arcsec, 6.4\arcsec, and 6.5\arcsec, respectively), and the astrometric precision of \textit{WISE} is typically $\lesssim 0.5$\arcsec\footnote{http://wise2.ipac.caltech.edu/docs/release/allwise/expsup/sec6\_4.html}. We obtained 10\arcmin\ $\times$ 10\arcmin\ images in 3.4, 4.6, and 12 \um\ from the \textit{WISE} image archives\footnote{http://irsa.ipac.caltech.edu/applications/wise/}. We used a setup of SExtractor, optimized for \textit{WISE} images\footnote{http://wise2.ipac.caltech.edu/staff/fmasci/SEx\_WPhot.html}, on each \textit{WISE} band separately. We compared the positions of sources extracted using SExtractor in the 3.4, 4.6, and 12 \um\ bands, using the closest extracted source to our expected source position based on the astrometry within the calibrated \textit{WISE} FITS header. We fit the core of the offset distributions (R.A. and Dec.) and found for the 3.4/4.6 \um\ offsets, each distribution had $\mu \approx 0$ and $\sigma \approx 1$\arcsec, and the 3.4/12 \um\ offsets had $\mu \approx 0$ and $\sigma \approx 5$\arcsec. Sources that had offsets $> 1$\arcsec\ in the 3.4/4.6 \um\ comparison failed extraction at 4.6 \um; these sources were found to have nearby bright sources as well as faint source detections, likely confusing SExtractor. Examples of objects that failed detection at 4.6 \um\ are shown in Figure~\ref{fig:sextractor1}. Figure~\ref{fig:wisedist} shows the results comparing the 3.4/12 \um\ source positions.

	To test if this scatter is intrinsically due to the precision of \textit{WISE} astrometry or a sign of contamination, we compared it against a subsample of SDSS DR10 quasars \citep[166,583 quasars;][]{paris:2014:a54}, extragalactic objects that are bright and remain stationary on the sky. Following step \#2 \& \#6 from \S\ref{sample}, we used only quasars that did not have another SDSS DR10 object within 6\arcsec. We also chose only quasars at high Galactic latitudes ($b\geq77^{\circ}$) to reduce contamination from stellar sources. Lastly, we chose quasars with a similar $W3$ SNR distribution to our candidates, which were found primarily between $3\leq \mathrm{SNR}_{W3} \leq 5$. Our final subset consisted of 304 quasars. We then followed the above procedure by obtaining 10\arcmin\ $\times$ 10\arcmin\ images in 3.4, 4.6, and 12 \um\ from \textit{WISE} and performed source extraction via SExtractor. As shown in Figure~\ref{fig:wisedist}, we found a similar offset distributions for our subsample of quasars compared to our candidate disk stars (a K-S test gives p-value$_\mathrm{R.A.} = 0.25$ and p-value$_\mathrm{Dec.} = 0.16$). The long tails of the distributions to higher offsets ($\gtrsim 10$\arcsec) mark objects where SExtractor failed to extract a source near our source position, and instead selected the closest extractable source. We include extracted position offsets in the online catalog.
		
	It is likely that SExtractor failed to extract our sources due to the low S/N threshold. We do not expect this to be a limitation to the \textit{WISE} pipeline, where the PSF is well characterized and the images are coadded, and the \textit{WISE} passive deblending method should be more reliable than SExtractor deblending. Further investigation is required into the \textit{WISE} pipeline to determine if source extraction from coadded images can be improved to include probabilities that each flux measurement is attributable to the same object in each band. The similarity of the quasar and disk candidate source position distributions likely implies a limitation to source extraction via SExtractor rather than a discrepancy in source position as all candidates were visually verified.

	To further investigate the possibility of chance alignments, we estimated the extragalactic source density that might contaminate our fields. Recently, \citet{yan:2013:55} estimated the number of \textit{WISE} extragalactic sources with S/N$_{W3}\geq3$, at Galactic latitudes $|b| > 20^\circ$, was 1,235 deg$^{-2}$. This background value was determined, down to a limiting flux density of $\sim$0.22 mJy ($W3\approx12.8$). Because the addition of the photospheric flux from a star plus the flux from chance alignment with an extragalactic source could create a high enough S/N flux to mimic an actual point-source detection, we expect our limiting flux density to be lower than the value found by \citet{yan:2013:55}. Our sample includes objects with magnitudes up to $W3=12.727$ ($\sim$0.236 mJy), however, we estimate that an extragalactic source that could potentially contaminate our sample could reach a magnitude of $W3 \approx 13$ ($\sim$0.183 mJy), which is the approximate \textit{WISE} detection limit for our sources (\S\ref{diskfrac}). 

	To estimate a new density of extragalactic sources down to $W3=13$, we sampled the AllWISE source catalog at 1,000 randomly chosen locations in the sky. We focused on Galactic latitudes $|b|\geq60^\circ$, since we expect fields at high Galactic latitudes to have the highest ratio of extragalactic to Galactic sources. We selected all sources (making no cuts on source morphology) within a 180\arcsec\ radius of our randomly chosen sky position, and with S/N$_{W3}\geq3$ and no contamination or confusion (\textsc{cc\_flags} = 0000). We repeated this process 15 times and found a source density of $976 \pm 14$ deg$^{-2}$, smaller than the value reported by \citet{yan:2013:55}, likely due to our higher Galactic latitude cut. 976 sources is potentially higher than the number of actual extragalactic sources since there are also stars in these fields. Considering this, we chose to be conservative and used the value of 976 sources as the number of contaminants we might find within a solid angle of 1 deg$^2$ centered about our candidate stars. 
	
\begin{figure}
 \plotone{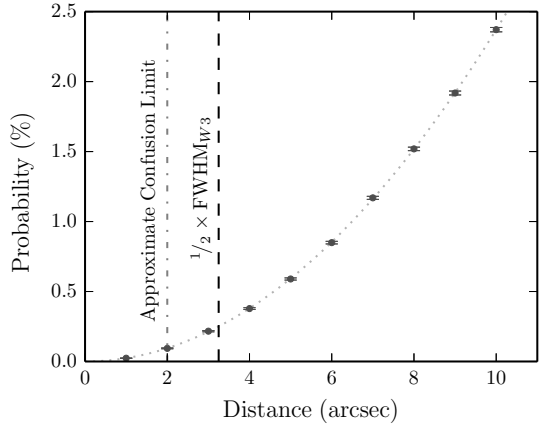}
\caption{The probability of chance alignment with an extragalactic source as a function of distance. Plotted are the computed probabilities with errors, assuming Poisson statistics, from our Monte Carlo method (circles), and the best-fit squared power law for our probabilities (dotted line). The resolution limit for $W3$ plotted (half of the FWHM, 3.25\arcsec; dashed line). The approximate confusion limit is shown as the dash-dotted line (see text). \label{fig:extragalactic}}
\end{figure}

	We simulated a circular patch of sky with $r =$ 2,031\arcsec\ (corresponding to a solid angle of 1 deg$^2$) and populated it with 976 sources randomly distributed within this circular area. Assuming our candidate star was centered on the circular area, we used a circular aperture centered on our stars position (the center of the circular area) and proceeded to count the number of randomly placed sources within our aperture, using an aperture radius from 1\arcsec--10\arcsec, in steps of 1\arcsec. We repeated this process $10^6$ times and computed probabilities assuming Poisson statistics for the uncertainties. These probabilities represent the likelihood of having an extragalactic source fall within some radius of our star, assuming extragalactic sources are randomly distributed on the sky. The results of our process are shown in Figure~\ref{fig:extragalactic}, where our probabilities follow a squared power law as expected for circular apertures. Assuming the confusion limit (i.e., the limit at which two objects cannot be distinguished from one) for the $W3$ beam is approximately half of the PSF FWHM (3.25\arcsec), then we can estimate the probability for chance alignment with an extragalactic source is $< 0.3$\%. This is a conservative estimate since a source at the confusion limit will appear significantly offset within the \textit{WISE} images, and may also appear as an extended source. If we assume a more likely confusion limit is $< 2$\arcsec, this leads to a probability of chance alignment $\lesssim 0.09\%$.
	
	For the computed probabilities to be meaningful, we assessed how many stars in the W11 catalog could have been observed if they had a similar IR excess to our disk candidates. We used a subsample of the W11 catalog that passed the selection criteria outlined in \S\ref{sample}, not including the \textit{WISE} S/N cuts. We required a magnitude limit to represent the minimal flux detectable by \textit{WISE} at 12 \um; since {\em WISE} sensitivity is highly dependent upon sky position due to the depth of coverage, we used \textit{WISE} $5\sigma$ sensitivity maps\footnote{http://wise2.ipac.caltech.edu/docs/release/allsky/expsup/sec6\_3a.html} to estimate the maximum detectable Vega magnitude for stars in the W11 catalog. Using the sensitivity maps, we estimated a detectable $W3$ Vega magnitude of $13 \pm 0.3$. The sensitivity maps are based on a noise model, and their flux densities were found to be slightly lower than those produced by external checks, therefore, we used the $1\sigma$ limit for our flux density cutoff (i.e., $W3=12.7$ or $\sim$0.24 mJy). For comparison, the minimum $W3$ flux density in our sample was 0.272 mJy. 
	
	Since we did not have reliable $W3$ values for the majority of the W11 catalog, we used this value to estimate a minimal detectable flux in another band for which we had information (i.e., $ugrizJHK_s$). We found that $J$-band magnitudes were good estimators for limiting fluxes due to the fact that the $J$-band is close to the peak of the SEDs for effective temperatures in the M-spectral type regime (2300--3800 K). We computed $J$-band magnitudes from synthetic photometry for the range $T_\ast=$ 2,300--3,800 K, solar metallicity, log$(g) =5$, using stellar radius estimates from \citet{reid:2005:}, at distances between $d=50$--1,600 pc in steps of 10 pc, representative of the distances within our sample. To determine an appropriate expected IR excess level, we used the average excess within our sample, found to be $\sim$10 times the photospheric level, and used this value to compute synthetic $W3$ magnitudes for each stellar model to find the corresponding $J$-band magnitude at the cutoff $W3$ magnitude of 12.7. On average, a $J$-band magnitude of $\sim$17 corresponded to a $W3$ magnitude of 12.7 for an excess 10 times the photospheric level. We implemented $J \le 17$ cutoff, and the resulting catalog contained 41,120 stars that could have been detected in $W3$ with SNR $\ge 3$ at the $5\sigma'$ level if they had an equivalent IR excess. 
	
	Therefore, for the 41,120 stars that could have a detectable excess, we expect $\lesssim 37$ false positives (0.09\% from above). We expect that our visual inspection should have removed many of these, and we will still retain a high number of true detections within our sample. At present, there are no known field M dwarfs ($\gtrsim 1$ Gyr) with IR excess detections. Higher resolution observations with \textit{JWST} in the future will make it possible to determine how many sources actually have IR interlopers. From our above results, we conclude that most of our observed excesses originate from the M dwarf source, rather than any contaminant.

\section{Results}\label{results}

\subsection{Modeling the Dust Population}\label{dust}

\begin{figure}
 \plotone{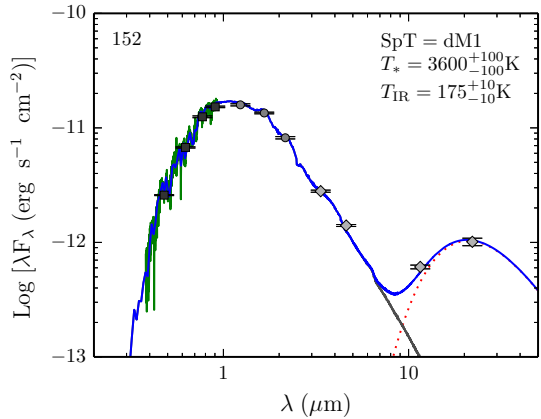}
\caption{A sample SED for one of the 17 stars with its IR excess fitted with a single temperature blackbody. Plotted are SDSS photometry 
(squares; $u$-band omitted due to large uncertainties for cool stars), 2MASS (circles), and \textit{WISE} (diamonds), 
along with SDSS spectra (green). The catalog number is listed in the upper left. The best-fit combined model is plotted in blue,
along with each of the model components (photosphere in gray, dust blackbody as dashed red line).
Source spectral type, model effective temperature, and blackbody temperature are listed in the top right hand corner of each panel. 
SEDs for the entire sample are available online. 
\label{fig:SED}
}
\end{figure}

	If we make the assumption that the excess flux is from a dust population radiating as a blackbody, we can examine the ratio of the dust and stellar luminosity. For our blackbody model fits, we attempted to fit all stars with S/N$_{W4} \geq 2$, excluding stars with \textit{WISE} $W4$ upper limits. We did not fit blackbodies to stars that had only 22 \um\ flux upper limits since blackbody fits using only 12 \um\ fluxes are extremely degenerate. We fit 17 of our candidate stars with a single temperature blackbody to model the observed IR excess. This was done using an iterative process where we fit blackbodies with temperatures from 50 to 1,000 K in steps of 5 K, that were added to the flux density from the stellar photosphere, and computed a $\chi^2$ minimization using all photometric bands. We followed a similar process to that outlined in \S\ref{methods}, except the newly computed blackbody flux density was normalized to the observed $W4$ flux density rather than the $K_s$-band. An example of a best-fit dust blackbody is shown in Figure~\ref{fig:SED}.

 	Fractional luminosities, $L_\mathrm{dust}/L_\ast$, can be estimated by taking a ratio of the integrated stellar flux and integrated thermal blackbody flux of the dust. This provides a lower limit to the flux produced by the dust population since there may be more FIR and submillimeter flux produced by colder dust grains. For sources with 12 \um\ excesses and 22 \um\ non-detections, we estimated the fractional IR excess with the assumption that $F_\mathrm{dust} \approx \nu F_\nu$, after subtracting from $F_{\nu}$ the expected stellar flux density contribution. For a thermal body peaking at this frequency, this approximation will typically be $\sim$74\% of the integrated thermal emission. This can be shown in the following approximation:
\begin{equation}
\frac{\lambda_\mathrm{max}F_{\lambda_\mathrm{max}}}{\int_0^\infty F_\lambda\, \mathrm{d}\lambda} = \frac{15h^4c^4}{b^4k^4\pi^4(e^{hc/bk}-1)}\approx 0.736,
\end{equation}
where $k$ is the Boltzmann constant, $c$ is the speed of light, $h$ is the Planck constant, and $b$ ($= 0.0036697$ m$\cdot$K) is the constant found from taking the derivative of $\lambda F_\lambda$ and solving for the peak value. 

	In the MIR, we can estimate the fractional luminosity of the dust to the star using the relationship between flux and luminosity, and the fact that disk emission peaks while the stellar flux will be in the Rayleigh-Jeans limit in this wavelength regime. The flux ratio between the dust population and host star becomes:
\begin{equation}
\frac{F_\mathrm{dust}}{F_\ast} = \frac{L_\mathrm{dust}}{L_\ast} \frac{hcT_\ast^3}{\lambda k T_\mathrm{dust}^4(e^{hc/\lambda kT_\mathrm{dust}}-1)}.
\end{equation}
The minimum fractional luminosity for our stars with only 12 \um\ excesses can be simplified assuming the peak emission is at 12 \um\ (corresponding to $T_\mathrm{dust} \approx 306$ K):
\begin{equation}
\frac{L_\mathrm{dust}}{L_\ast}\mathrm{(minimum)} = 10^{-2} \left(\frac{3304\, \mathrm{K}}{T_\ast} \right)^3 \frac{F_\mathrm{12,\, dust}}{F_\mathrm{12,\, \ast}}.
\end{equation}
Fractional luminosities are listed in Table~\ref{tbl:modelparams}. 

	We find that many of our stars exhibit large fractional luminosities ($\sim$$10^{-2}$--$10^{-1}$), approximately five orders of magnitude larger than estimates of the Kuiper belt's luminosity relative to the Sun \citep{beichman:2005:1160}. This is also larger than values typically associated with debris disks \citep[$L_\mathrm{dust} / L_\ast \sim 10^{-3}$;][]{rebull:2008:1484}. \citet{gautier:2008:813} observed similar fractional luminosities for low-mass stars in $\eta$ Chamaeleontis, a young \citep[4--15 Myr;][]{mamajek:2000:356,luhman:2004:816,lyo:2004:246} stellar association. Fractional luminosities $> 10^{-2}$ have been observed around both classical T Tauri stars (CTTSs) and weak-line T Tauri stars (WTTS) \citep{padgett:2006:1283,cieza:2007:308}, however, the majority of stars in our sample exhibit no H$\alpha$ emission, and are therefore likely older than 1 Gyr. We will discuss the implications of this in \S\ref{discussion}.

	We can further characterize the dust population if we assume it exists in radiative equilibrium with the host star. For a dust grain in radiative equilibrium with its host star, the distance from the host star is given by:
\begin{equation}
D=\frac{1}{2}\left(\frac{T_\ast}{T_\mathrm{gr}}\right)^2 R_\ast,
\end{equation}
\citep{ jura:1998:897, chen:2005:1372}. Here $T_\ast$ and $T_\mathrm{gr}$ are the stellar effective temperature and dust grain temperature, and $R_\ast$ is the stellar radius. For stars without 22 \um\ excesses we assumed $T_\mathrm{gr} \approx 306$ K, the radiative temperature of a blackbody at 12 \um. The average distance was calculated to be $\sim$0.16 AU for the combined sample, within the snow line for planets around M dwarfs \citep[$\sim$0.3 AU;][]{ogihara:2009:824}. Minimum distances are listed in Table~\ref{tbl:modelparams}.

	Fractional IR luminosities can be used to estimate a dust mass if we make a few assumptions. First, we assume an average grain radius, $a$, and density, $\rho_s$. For reasons we will discuss in \S{\ref{discussion}}, we assume the IR excesses we are viewing are due to planetary collisions rather than primordial dust content; then, using the results of \citet{weinberger:2011:72}, we assumed an average minimum grain size of $\langle a\rangle=0.5$ \um, noting that particles can have $a > 10$ \um. Grain size plays an important role in the observations, since larger grains, which radiate more efficiently, will show similar temperatures at closer orbital distances to smaller grains at larger orbital distances. Therefore, we used 0.5 \um\ as a lower limit similar to \citet{weinberger:2011:72}, noting that larger grains, which put dust populations closer to their host stars, do not change the results of our discussion (\S\ref{discussion}). We assumed a typical grain density of $\rho_s = 2.5$ g cm$^{-3}$ \citep{pollack:1994:615}. If we assume the dust is in a thin shell, orbiting a distance $D$ from the host star, with a particulate radius $a$ and density $\rho_s$, and a cross-section equal to the physical cross section of a spherical grain, we can define the dust mass as:
\begin{equation}
M_d\geq\frac{16}{3}\pi\frac{L_\mathrm{dust}}{L_\ast}\rho_s\langle a\rangle D_\mathrm{min}^2,
\end{equation}
\citep{chen:2005:1372}. Our computed dust masses are listed in Table~\ref{tbl:modelparams}. A few of our stars have dust masses on the order of AU Mic's \citep[$10^{-4} M_\mathrm{moon}$;][]{chen:2005:1372}. However, a typical disk mass in our sample is comparable to, if not slightly larger than, those observed for debris disks ($10^{-5} M_\mathrm{moon}$). IR spectra of these systems will help to further characterize and constrain the crystallinity of the dust content, similar to the methods used in \citet{weinberger:2011:72}. In particular, IR spectra may be useful to distinguish dust produced from planetary collisions within the terrestrial zone versus primordial dust. This will be discussed further in \S\ref{discussion}.

\subsection{Spectral Signatures of Youth}\label{youth}

	Low-mass stars not only have extremely long main sequence lifetimes, they have long pre-main sequence lifetimes, from 100s of Myrs to over a Gyr dependent on mass \citep{baraffe:2002:563}. Disk lifetimes around low-mass stars are not well constrained; some theories predict that low-mass stars can retain their disks for periods much longer than that for solar-type stars \citep[$\sim$10 Myr;][]{plavchan:2009:1068}. For comparison, the 12 Myr AU Mic \citep[dM1e;][]{wilner:2012:l27} has shown excess at both 22 \um\ (A12) and 70 \um\ \citep{plavchan:2009:1068}. Therefore, youth is important to quantify to help inform the timescale for disk dispersal.
	
	Most of our disk candidates appear to be field stars, not affiliated with any known star-forming region or young stellar association. However, there remains the possibility that some of the stars in our sample are members of unknown young stellar associations or moving groups, or unidentified members of known associations. To test for youth within our sample, we looked at tracers of surface gravity, magnetic activity, and lithium absorption. Youth was quantified through an index for each tracer measured. These indices are listed in Table~\ref{tbl:spec}.

\begin{deluxetable*}{lcccccc}
\tabletypesize{\scriptsize}
\tablecolumns{7}
\tablewidth{0pt}
\tablecaption{Spectral Measurements\label{tbl:spec}}
\tablehead{
\colhead{Catalog} & \colhead{H$\alpha$ EW} & \colhead{K \textsc{i} EW\tablenotemark{a}} & \colhead{Na \textsc{i} EW\tablenotemark{b}} & \colhead{CaH 3\tablenotemark{c}} & \colhead{TiO 5\tablenotemark{c}} & \colhead{Youth}\\
\colhead{Number} & \colhead{(\AA)} & \colhead{(\AA)} & \colhead{(\AA)} & \colhead{} & \colhead{} & \colhead{Index\tablenotemark{d}}
}
\startdata
40.......... & -2.78 $\pm$ 0.22 & 2.07 $\pm$ 0.11 & 1.71 $\pm$ 0.09 & 0.81 $\pm$ 0.01 & 0.61 $\pm$ 0.01 & 1010 \\
42.......... & -5.09 $\pm$ 0.23 & 2.13 $\pm$ 0.10 & 2.24 $\pm$ 0.07 & 0.77 $\pm$ 0.01 & 0.56 $\pm$ 0.01 & 1110 \\
62.......... & -6.82 $\pm$ 1.03 & 7.28 $\pm$ 0.74 & 5.91 $\pm$ 0.36 & 0.66 $\pm$ 0.03 & 0.33 $\pm$ 0.03 & 1100 \\
78.......... & -11.21 $\pm$ 1.07 & 8.72 $\pm$ 1.13 & 5.24 $\pm$ 0.34 & 0.56 $\pm$ 0.03 & 0.29 $\pm$ 0.03 & 1000 \\
108........ & 0.03 $\pm$ 0.07 & 1.99 $\pm$ 0.13 & 2.31 $\pm$ 0.13 & 0.85 $\pm$ 0.01 & 0.74 $\pm$ 0.01 & 0000 \\
110........ & -0.05 $\pm$ 0.21 & 2.96 $\pm$ 0.35 & 4.06 $\pm$ 0.29 & 0.74 $\pm$ 0.03 & 0.48 $\pm$ 0.03 & 0000 \\
115........ & 0.55 $\pm$ 0.11 & 1.71 $\pm$ 0.21 & 2.33 $\pm$ 0.21 & 0.85 $\pm$ 0.02 & 0.76 $\pm$ 0.03 & 0000 \\
126........ & 0.21 $\pm$ 0.15 & 3.03 $\pm$ 0.31 & 1.85 $\pm$ 0.26 & 0.85 $\pm$ 0.03 & 0.68 $\pm$ 0.03 & 0000 \\
140........ & -84.02 $\pm$ 16.83 & 4.35 $\pm$ 0.38 & 3.39 $\pm$ 0.14 & 0.70 $\pm$ 0.01 & 0.41 $\pm$ 0.01 & 1000 \\
142........ & 0.36 $\pm$ 0.08 & 1.24 $\pm$ 0.18 & 1.61 $\pm$ 0.15 & 0.88 $\pm$ 0.02 & 0.80 $\pm$ 0.02 & 0000 \\
152........ & 0.14 $\pm$ 0.04 & 1.93 $\pm$ 0.09 & 1.56 $\pm$ 0.08 & 0.85 $\pm$ 0.01 & 0.64 $\pm$ 0.01 & 0001 \\
156........ & 0.30 $\pm$ 0.35 & 2.05 $\pm$ 0.59 & ... & 0.77 $\pm$ 0.05 & 0.69 $\pm$ 0.06 & 0000 \\
164........ & 0.30 $\pm$ 0.08 & 1.98 $\pm$ 0.16 & 1.62 $\pm$ 0.16 & 0.89 $\pm$ 0.01 & 0.81 $\pm$ 0.02 & 0000 
\enddata
\tablecomments{Values listed only for stars with SNR$_{W4} \geq 3$. Values for the combined sample are available online. Negative EWs indicate emission.}
\tablenotetext{a}{K \textsc{i} doublet (7665 \& 7699 \AA).}
\tablenotetext{b}{Na \textsc{i} doublet (8183 \& 8195 \AA).}
\tablenotetext{c}{Defined in \citet{reid:1995:1838}.}
\tablenotetext{d}{Youth index: H$\alpha$ emission, low surface gravity in at least 2 of the 3 tracers (\S\ref{gravity}), lithium absorption, and UV activity.}
\end{deluxetable*}

\subsubsection{Surface Gravity}\label{gravity}

	Pre-main sequence stars that are still contracting onto the main sequence will exhibit weaker surface gravity relative to stars on the main sequence. Several atomic/molecular bands have been identified as reliable tracers of surface gravity. To estimate surface gravities for our sample, we analyzed the atomic alkali lines of Na \textsc{i} (8183 and 8195 \AA) and K \textsc{i} (7665 and 7699 \AA) identified in \citet{slesnick:2006:3016}, and the CaH 3 molecular absorption band from \citet{reid:1995:1838}. Other studies of nearby, young low-mass stars have used these same molecular features as proxies for youth \citep[e.g.,][]{shkolnik:2009:649,shkolnik:2011:6,shkolnik:2012:56,schlieder:2012:114}. 
	
	One caveat to using molecular bands as a diagnostic for surface gravity is their dependence on stellar metallicity. We can interpret the findings of \citet[][]{woolf:2006:218} to surmise that higher metallicity will increase both the Na \textsc{i} and K \textsc{i} equivalent width (EW) measurements. Therefore, a star with low surface gravity and high metallicity may appear to have high surface gravity. Metallicities are difficult to measure for low-mass stars due to their large abundance of molecular features and incomplete line lists. Many studies attempted to find metallicity tracers using methods both photometric \citep{bonfils:2005:635,johnson:2009:933,schlaufman:2010:a105} and spectroscopic \citep{lepine:2007:1235,rojas-ayala:2010:528,terrien:2012:l38,mann:2013:52,newton:2014:20}. We employed the methods outlined in \citet{mann:2013:52} to determine metallicity using the SDSS spectra (\S\ref{metallicity}). We find that the majority of our stars appear to have solar metallicity, and should not affect our gravity sensitive spectroscopic features excessively. Considering this we used a comparative approach for estimating surface gravities.
	
	The CaH 3 index was computed following the method outlined in \citet{reid:1995:1838}. Line measurements were made by integrating over a specific width centered on each line, and subtracting off the mean flux calculated from two adjacent continuum regions; regions are listed in Table~\ref{tbl:lines}. EWs were computed for each line by dividing the integrated line flux by the mean continuum value. Formal EW uncertainties for each line were also computed. We integrated the Na \textsc{i} doublet over a single region, and summed two separate regions for the K \textsc{i} doublet.
	
\begin{deluxetable}{ccccc}
\tabletypesize{\scriptsize}
\tablecolumns{5}
\tablewidth{0pt}
\tablecaption{Line Measurement Regions\label{tbl:lines}}
\tablehead{
\colhead{Line} & \colhead{Central $\lambda$} & \colhead{Width} & \colhead{Continuum A} & \colhead{Continuum B} \\
\colhead{} & \colhead{(\AA)} & \colhead{(\AA)} & \colhead{(\AA)} & \colhead{(\AA)} 
}
\startdata
Na \textsc{i} & 8189 & 22 & 8149--8169 & 8236--8258\\
K \textsc{i} a & 7665 &16 & 7651--7661 & 7680--7690\\
K \textsc{i} b & 7699 &16 & 7680--7690 & 7710--7720
\enddata
\tablecomments{All of the wavelengths are given in vacuum units.}
\end{deluxetable}

\begin{figure}

 \plotone{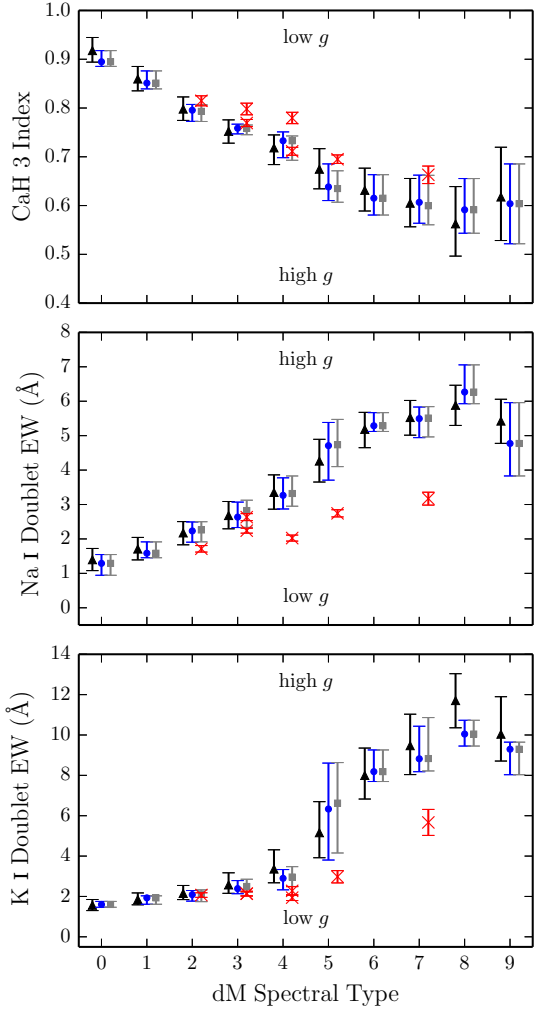}
\caption{Log($g$) tracers for the combined sample (blue circles), the combined sample without the Orion candidates (gray squares), Orion candidates (red crosses), and the W11 catalog (black triangles). All data points represent the median value with error bars representing the $1^\mathrm{st}$ and $3^\mathrm{rd}$ quartiles for each spectral type, except for the Orion candidates which show actual measurements and uncertainties. As an ensemble, the combined sample does not appear to have characteristically lower surface gravities compared to the W11 catalog, however, the Orion candidate stars appear to have low-surface gravity relative to the W11 values (see Table~\ref{tbl:surfacegravity}). 
\label{fig:logg}}
\end{figure}

	We expect the majority of stars in the W11 catalog to be older field dwarfs. Therefore, to determine low surface gravity members of our sample, we compared gravity tracers for our sample relative to the W11 catalog. We grouped stars by spectral type and determined quartiles for each gravity tracer and spectral type bin. We plot our surface gravity tracers in Figure~\ref{fig:logg}, compared to the W11 catalog. We expect low surface gravity members of the W11 catalog to appear as outliers in each gravity tracer within each spectral type bin. The majority of the stars in our sample fall within the intrinsic scatter of the W11 catalog (between the $1^\mathrm{st}$ and $3^\mathrm{rd}$ quartiles), indicating that we are observing an older stellar population rather than pre-main sequence stars. Each star that shows lower surface gravity relative to the W11 quartiles (i.e., below the $1^\mathrm{st}$ quartile for Na \textsc{i} and K \textsc{i}, or in the $4^\mathrm{th}$ quartile for CaH 3) in at least two of the three tracers has been marked in Table~\ref{tbl:spec}. Overall, the combined sample does not appear to have significantly lower surface gravities over the W11 catalog, however, the Orion candidates either fall below or straddle the lower end of the W11 catalog quartiles for each surface gravity tracer, implying that it is likely a younger stellar population. 
	
	For our sample we computed the percentages of stars with surface gravities lower than the W11 $3^\mathrm{rd}$ quartiles, and lower than the W11 medians, respectively; values are listed in Table~\ref{tbl:surfacegravity}. The combined sample has only a small percentage of low-surface gravity stars ($<15$\%). The large percentage of stars with low-surface gravity towards Orion indicates that these stars are probably members of the star-forming region rather than field stars along the LOS. These stars will be discussed further in \S\ref{orion}.

\begin{deluxetable}{ccc}
\tabletypesize{\scriptsize}
\tablecolumns{3}
\tablewidth{0pt}
\tablecaption{Surface Gravity Statistics\label{tbl:surfacegravity}}
\tablehead{
\colhead{} & \multicolumn{2}{c}{\# of Stars with log($g$) $<$} \\ 
\colhead{Sample} & \colhead{W11 3$^\mathrm{rd}$ Quartile} & \colhead{W11 Median} 
}
\startdata
Combined & 17.7\% & 47.4\% \\
Combined w/o Orion & 14.9\% & 45.2\% 
\enddata
\end{deluxetable}

\subsubsection{Lithium Absorption}\label{lithium}

	Young pre-main sequence M dwarfs that are still contracting onto the main sequence rapidly destroy their natal lithium content as their interiors reach the required temperature ($\sim$$2.5\times10^6$ K). Once they have contracted onto the main sequence, an early-type M dwarf will deplete its lithium by a factor of 2 in less than 10 Myr, whereas a late-type M dwarf may show absorption at $\sim$100 Myr \citep{cargile:2010:l111}. In addition, fully convective stars \citep[$M \lesssim 0.35 M_\odot$;][]{chabrier:1997:1039} that exhibit episodic accretion can increase the rate of lithium depletion by increasing central temperatures, with complete lithium depletion occurring between 10 to a few 100 Myr in models \citep{baraffe:2010:a44}. Therefore, measuring the lithium feature (Li \textsc{i} 6708 \AA) provides a strong youth tracer. 
	
	Due to the low S/N around the Li \textsc{i} feature, and the strong TiO bands in close proximity, a direct measurement of this feature was not attempted. Instead, we decided to compare our spectra to the template spectra from \citet{bochanski:2007:531}. \citet{bochanski:2007:531} created template spectra through co-addition of $\sim$ 4,300 high S/N SDSS spectra of field M dwarfs. These template spectra represent a statistically older population of M dwarfs, with little to no lithium absorption in their spectra. Comparing these template spectra to our sample SDSS spectra, we can identify significant lithium depletion in comparison to the template to identify stars that are younger than average field stars. To determine the significance of the absorption feature, we used a method similar to that in \citet{cargile:2010:l111}, using a moving integrated residual. The moving integrated residuals were determined by integrating over the residuals from the template spectrum minus the SDSS spectrum in a moving 5 \AA\ window. This window was moved in steps of 2 \AA. Using the moving window, small differences between the template and source spectrum are averaged out, providing a better measure of the deviations between the two. Standard deviations are derived from the R.M.S. of the residual outside a 10 \AA\ window that is centered at 6708 \AA. For further details on this method see \citet{cargile:2010:l111}. 
	
\begin{figure}
 \plotone{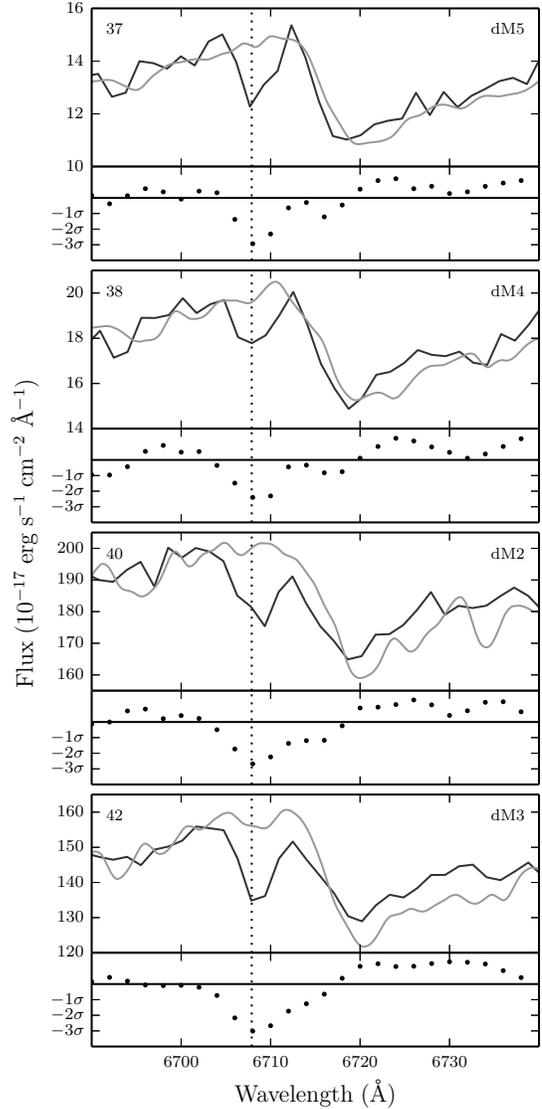}
\caption{Spectral profiles (dark gray line) for stars in the combined sample showing lithium absorption at 6708 \AA\ (dotted line). 
SDSS template spectra from \citet{bochanski:2007:531} are plotted for comparison
(light gray line). Below each spectra are the moving integrated residuals showing Li absorption. 
The template spectra represent older field dwarfs ($\gtrsim$1 Gyr) with little to no absorption.
Catalog number and spectral type are listed in the top left and right corner, respectively. 
\label{fig:lithium}}
\end{figure}

	We applied the \citet{cargile:2010:l111} method on every star in our sample and found \lithiumstars\ of the \totalsample\ had a detectable amount of lithium. The method can be seen in Figure~\ref{fig:lithium}, where we display the \lithiumstars\ stars in our sample showing significant lithium absorption relative to the template spectra. These stars are marked in Table~\ref{tbl:spec} and all four are found in the Orion footprint. Due to the assumed ages of field stars within the W11 catalog ($\gtrsim1$ Gyr), we would expect none of these stars to show detectable amounts of lithium if they are members of the field. Although we make no attempt to assign ages to these stars, it is probable that they are not members of the field and are younger stars associated with the Orion OB1 association.
		
\subsubsection{H$\alpha$ Emission}\label{halpha}

	Over the past few decades, numerous studies have linked magnetic activity (traced through hydrogen emission) to stellar age \citep[e.g.,][]{wilson:1963:832, skumanich:1972:565, eggen:1990:166, soderblom:1991:722, hawley:1996:2799, hawley:2000:252}. W08 found that for M dwarfs with spectral types earlier than dM4, the average activity lifetime was less than 2.5 Gyrs, increasing to as long as 8 Gyrs for stars with spectral types later than dM4. Although the assumed primordial disk dispersal time is far shorter than the activity lifetime for any M dwarf (10s of Myrs versus Gyrs), H$\alpha$ provides another diagnostic that we can use to differentiate a statistically younger population from an older one, particularly for stars with spectral types earlier than dM4. Hydrogen transitions have also been observed as a sign of accretion, with an increase in line width measurements over typical magnetically active stars \citep{white:2003:1109}.
	
	Using the RVs computed in this study, we remeasured the hydrogen Balmer transitions following the methods outlined in W11 to analyze the magnetic activity within our sample. Stars were flagged as active (\textsc{actha} $=1$) if they met the following criteria: the H$\alpha$ equivalent width (EW) $<0.75$ \AA, the S/N in the continuum was greater than three, the EW value was three times the uncertainty, and the height of the spectral line was larger than three times the noise in the continuum. The small size of our sample allows us to visually inspect each spectrum to make sure the activity flag corresponded to a real signal, and did not potentially miss stars that exhibited H$\alpha$ emission. We found all stars flagged as active to show some amount of H$\alpha$ emission visible in the SDSS spectrum, with none of the stars flagged as inactive showing emission. H$\alpha$ EW measurements are listed and active candidates are marked in Table~\ref{tbl:spec}. 
	 
\begin{figure}
 \plotone{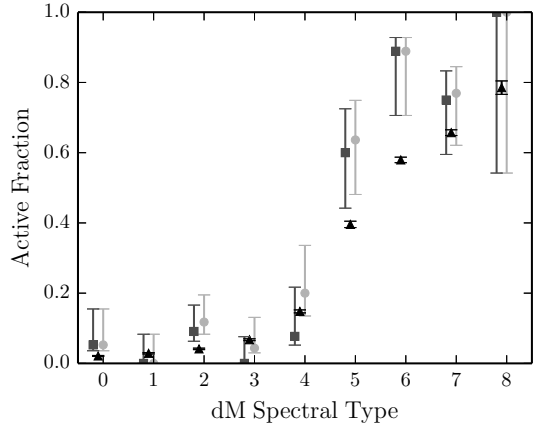}
\caption{Activity fractions for stars from the combined sample (circles), the combined sample without the Orion candidates (squares), and the W11 catalog (triangles). The error bars represent binomial distribution errors. Without the Orion candidates, the combined sample typically has lower activity fractions than the W11 catalog, implying the Orion candidates are mostly younger, active stars. As an ensemble, we do not see significantly different activity fractions between the W11 catalog and our combined sample.}\label{fig:activity}
\end{figure}

	To compare the level of activity in our sample to a relatively unbiased sample of field stars, we computed activity fractions ($N_\mathrm{active}/N_\mathrm{total}$) for each spectral type bin and compared them to the W11 values. We define stars as active using the same criteria as W11. For the total number of stars, we only used stars that were defined as active or inactive (using the W11 criteria), omitting stars that did not have high enough S/N to categorize their activity status. Figure~\ref{fig:activity} shows the activity fractions for stellar samples from this study versus the W11 catalog. The number of stars that went into each bin are listed in Table~\ref{tbl:activity}. Although our small sample size makes it difficult to draw strong conclusions, we can see that activity fractions are not significantly different between our sample and the W11 catalog. If we remove the Orion candidates from our combined sample, activity fractions drop for each spectral-type bin that contains Orion candidates. This suggests that the majority of ``young" stars in our sample are the Orion candidates, giving additional weight to the possibility that those stars are indeed members of the Orion OB1 association. Considering the long activity lifetimes for low-mass stars, the majority of our combined sample do not show characteristics of a young stellar population ($< 100$ Myr). If our stellar sample is older than $\sim$1 Gyr, this suggests that either the IR excesses we observe do not result from primordial circumstellar material, or that the disk dispersal time for low-mass stars can last for much longer than current theories and empirical data predict (10s of Myrs). Further discussion of interpretations for these IR excesses will ensue in \S\ref{discussion}.

\begin{deluxetable*}{lccccccccccc}
\tabletypesize{\scriptsize}
\tablecolumns{12}
\tablewidth{0pt}
\tablecaption{Activity Indicators\label{tbl:activity}}
\tablehead{
&&& \colhead{Active} &&&& \colhead{Active} &&&& \colhead{Active} \\ 
\colhead{SpT} & \colhead{$N_\mathrm{active}$} & \colhead{$N_\mathrm{inactive}$} & \colhead{Fraction} && \colhead{$N_\mathrm{active}$} & \colhead{$N_\mathrm{inactive}$} & \colhead{Fraction} && \colhead{$N_\mathrm{active}$} & \colhead{$N_\mathrm{inactive}$} & \colhead{Fraction}
}
\startdata
& \multicolumn{3}{c}{Combined Sample} && \multicolumn{3}{c}{Combined Sample w/o Orion} && \multicolumn{3}{c}{\citet{west:2011:97} Catalog}\vspace{0.1cm}\\
\cline{2-4}\cline{6-8}\cline{10-12} 
&&&&&&&&&&&\vspace{-0.2cm}\\
M0 & 1 & 18 & $0.05^{+0.10}_{-0.02}$ && 1 & 18 & $0.05^{+0.10}_{-0.02}$ && 207 & 9573 & $0.02^{+0.01}_{-0.01}$ \\
M1 & 0 & 20 & $0.00^{+0.08}_{-0.01}$ && 0 & 20 & $0.00^{+0.08}_{-0.00}$ && 226 & 7649 & $0.03^{+0.01}_{-0.01}$ \\
M2 & 4 & 30 & $0.12^{+0.08}_{-0.03}$ && 3 & 30 & $0.09^{+0.08}_{-0.03}$ && 366 & 8496 & $0.04^{+0.01}_{-0.01}$ \\
M3 & 1 & 22 & $0.04^{+0.09}_{-0.01}$ && 0 & 22 & $0.00^{+0.08}_{-0.00}$ && 623 & 8630 & $0.07^{+0.01}_{-0.01}$ \\
M4 & 3 & 12 & $0.20^{+0.14}_{-0.07}$ && 1 & 12 & $0.08^{+0.14}_{-0.02}$ && 1040 & 5983 & $0.15^{+0.01}_{-0.01}$ \\
M5 & 7 & 4 & $0.64^{+0.11}_{-0.16}$ && 6 & 4 & $0.60^{+0.12}_{-0.16}$ && 1143 & 1745 & $0.40^{+0.01}_{-0.01}$ \\
M6 & 8 & 1 & $0.89^{+0.04}_{-0.18}$ && 8 & 1 & $0.89^{+0.04}_{-0.18}$ && 2288 & 1661 & $0.58^{+0.01}_{-0.01}$ \\
M7 & 10 & 3 & $0.77^{+0.08}_{-0.15}$ && 9 & 3 & $0.75^{+0.08}_{-0.16}$ && 2236 & 1167 & $0.66^{+0.01}_{-0.01}$ \\
M8 & 2 & 0 & $1.00^{+0.01}_{-0.46}$ && 2 & 0 & $1.00^{+0.01}_{-0.46}$ && 362 & 99 & $0.79^{+0.02}_{-0.02}$ 
\enddata
\end{deluxetable*}

\subsubsection{Ultraviolet Flux}\label{uv}

	Heating of the chromosphere also produces ultraviolet (UV) emission through a number of transitions \citep{linsky:2001:247,findeisen:2010:1338}, which provides another tracer for stellar activity, and hence youth. Included in the UV emission spectrum are the Mg \textsc{ii} resonance lines, a host of Fe \textsc{ii} lines, and other species including: Ly$\alpha$, C \textsc{ii}, Al \textsc{ii}, C \textsc{iii}, Si \textsc{iii}, Si \textsc{iv}, and He \textsc{ii} \citep{walkowicz:2008:593,france:2012:l32}. \citet{shkolnik:2011:6} determined criteria to separate older populations from young populations of M dwarfs using the \textit{Galaxy Evolution Explorer} \citep[\textit{GALEX};][]{martin:2005:l1} near ultra-violet (NUV, 1750--2750 \AA) and far ultra-violet (FUV, 1350--1750 \AA) fluxes. 
	
	Using a cross-matched catalog between \textit{GALEX} Data Release 6 sources and the W11 catalog (Jones \& West 2014, in preparation), we found five matches to our sample that had flux measurements in the NUV, FUV, or both. The fraction of our sample (2.9\%) matched to \textit{GALEX} sources that had either an NUV or FUV detection is comparable to the fraction of sources Jones \& West (2014, in preparation) were able to match to the entire W11 catalog (1.4\%). Due to the distances to stars in the W11 catalog ($d\gtrsim100$ pc), UV emission from \emph{inactive} M dwarfs should be undetectable (Jones \& West 2014, in preparation). None of the matched \textit{GALEX} objects were associated with any of our Orion stars. 
	
	To determine M dwarfs showing significant UV flux as a tracer for stellar activity, \citet{shkolnik:2011:6} compared \textit{GALEX} fluxes to 2MASS $J$-band fluxes to separate young, active stars from quiescent, older stars and white dwarf/M dwarf close binaries. One of the stars in our combined sample had both NUV and FUV fluxes, and significant fractional UV fluxes ($F_\mathrm{NUV}/F_{J} \approx F_\mathrm{FUV}/F_{J} > 10^{-4}$), and is flagged in Table~\ref{tbl:params}. All five stars that showed NUV flux, independent of whether they had a FUV detection, show significant NUV excess ($F_\mathrm{NUV}/F_{J} \gtrsim 10^{-4}$). We would expect stars showing strong UV flux to exhibit other chromospheric heating signs, however, only one of the five stars exhibit H$\alpha$ emission or any other youth tracer. \citet{shkolnik:2011:6} estimate the probability of flaring M dwarfs contributing to \textit{GALEX} source counts to be $<3$\%, therefore, although unlikely, it is possible \textit{GALEX} observed these stars during flare activity.

\subsection{Kinematics}\label{kinematics}

	Stellar kinematics can provide additional insight into the age of a stellar population. To closely examine the kinematics of our stars, we used the USNO-B/SDSS proper motions \citep{munn:2004:3034,munn:2008:895}, radial velocities computed in this study, and photometric parallax distances to compute $UVW$ space motions. $UVW$ velocities are in a right-handed Cartesian coordinate system, representing motion towards the Galactic center, motion in the direction of the Sun's motion, and motion northwards away from the Galactic plane, respectively. For stars at appreciable distances, $UVW$ velocities may not be appropriate tracers of Galactic motion, however, the vast majority of our stars are within the distance limits used by other studies of stellar kinematics \citep[$d <$ 1,600 pc, $|Z| <$ 1,000 pc;][]{bochanski:2007:2418}. Younger stars will typically have small $UVW$ motions with respect to the local standard of rest (LSR), with older stellar populations lagging the LSR along the direction of galactic rotation ($V$ component), a phenomenon known as ``asymmetric drift" \citep{stromberg:1922:265,stromberg:1925:363}. 
	
\begin{figure*}
 \plotone{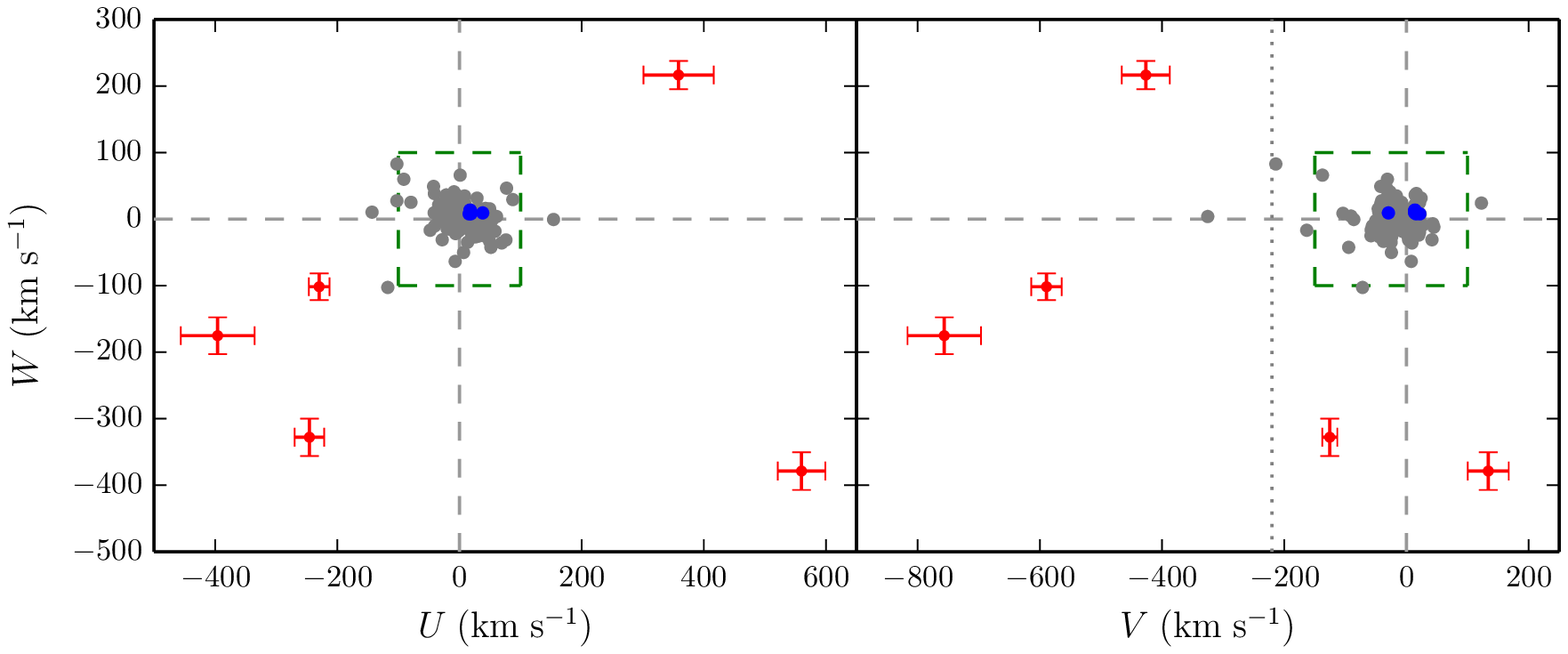}
 \plotone{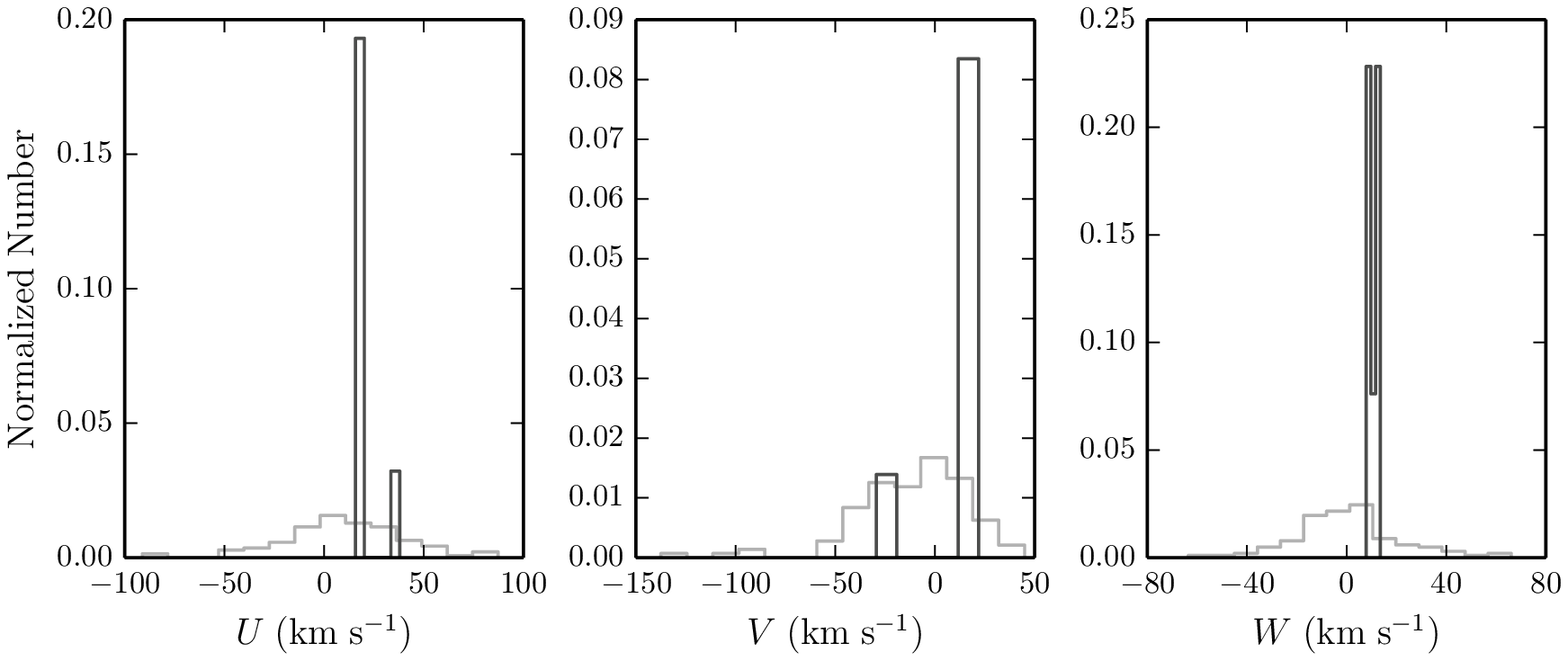}
\caption{Stellar kinematics for stars with good proper motions (\textsc{goodpm} $= 1$). Top: The combined sample (gray points) with high velocity stars marked ($v_\mathrm{tot} > 400$ km s$^{-1}$; red crosses). Orion stars are plotted as blue points. The majority of stars appear to have kinematics typical of the disk population, with a few exhibiting lagging $V$ velocity components, expected from older M dwarfs (see text for details). The dotted line is drawn at $V = -220$ km s$^{-1}$, separating stars which are moving along the direction of Galactic rotation from stars moving counter to it. Bottom: Normalized velocity distributions for stars that fall within the green dashed line for our combined sample minus the Orion candidates (light gray) and Orion candidates (dark gray). The Orion candidates show far less dispersion in their $UVW$ velocities, indicating a younger dynamical population. The tail of the distribution shown in the Orion candidates indicates there may be a few field stars along our line of sight. The largest source of error for the high velocity stars are the photometric distance errors, which may be $\sim$20--30\% of the distance measurement.\label{fig:uvw}}
\end{figure*}

	$UVW$ velocities and uncertainties were computed following the method of \citet{johnson:1987:864} by way of a Python version of the IDL procedure \texttt{gal\_uvw.pro}. These values were then corrected for the solar motion \citep[$U = 11.1 \textrm{ km s}^{-1}$, $V = 12.24 \textrm{ km s}^{-1}$, $W = 7.25 \textrm{ km s}^{-1}$;][]{schonrich:2010:1829} with respect to the LSR. The kinematics of our samples are shown in Figure~\ref{fig:uvw}. As an ensemble, our stars appear to have kinematics typical of the disk population \citep{leggett:1992:351}. The Orion candidates appear to be distributed about the origin of velocity space, indicating a younger stellar population. There also appears to be five stars with velocities on the order of, or larger than, the escape velocity of the Galaxy \citep[$\sim$400--500 km s$^{-1}$;][]{piffl:2014:a91}. 
	
	Older stellar populations will have been dynamically heated over time, causing a larger velocity dispersion as a population. As is shown in Figure~\ref{fig:uvw}, our Orion candidates have narrow $UVW$ velocity distributions compared to our combined sample, indicating a younger, less dynamically heated stellar population. We can also see the skew of the $V$ distribution for our combined sample to more negative values, indicative of asymmetric drift. There are a few outliers in our Orion velocity distributions, most likely due to field stars along our line of sight. We do not perform an in-depth investigation on the candidacy of these stars here, although one is warranted. The computed average and dispersion of the $UVW$ components for our combined sample without the Orion candidates, as listed in Table~\ref{tbl:uvwsamp}, are similar to other studies of late-type stars \citep{fuchs:2009:4149}. We acknowledge that there is a selection bias due to the fact that we cannot probe the motions of stars in the distant Galaxy, where stellar motions are different due to the age of those stellar populations, however, this should not significantly affect our analysis of the Orion candidates.
	
\begin{deluxetable*}{ccccccc}
\tabletypesize{\footnotesize}
\tablecolumns{7}
\tablewidth{0pt}
\tablecaption{Velocity Components\label{tbl:uvwsamp}}
\tablehead{
\colhead{Sample} & \colhead{$\langle U\rangle$} & \colhead{$\sigma_{\langle U\rangle}$} & \colhead{$\langle V\rangle$} & \colhead{$\sigma_{\langle V\rangle}$} &
\colhead{$\langle W\rangle$} & \colhead{$\sigma_{\langle W\rangle}$} 
}
\startdata
\vspace{-0.15cm}Combined w/o & & & & & &\\ \vspace{-0.15cm}
 & $10.19 \pm 2.83$ & $29.70$ & $-11.98 \pm 2.78$ & $29.12$ & $0.59 \pm 2.03$ & $21.31$ \\
Orion Candidates & & & & & & \vspace{0.15cm}\\ 
Orion Candidates & $20.52 \pm 2.71$ & $7.18$ & $8.89 \pm 6.01$ & $15.90$ & $10.64 \pm 0.84$ & $2.23$ 
\enddata
\end{deluxetable*}
	
	Five of our stars showed high velocity kinematics, usually associated with ``hypervelocity" or runaway stars. The largest component of motion for these stars is tangential to our line of sight, therefore, we needed to make sure our proper motions were reliable. All five high proper motion stars passed the quality cuts defined in \citet[][{\sc sigRA} $<525$, {\sc sigDEC} $<525$]{kilic:2006:582}, which were an extension of the cuts defined by \citet{munn:2004:3034}. They were also found in multiple proper motion catalogs, and values are listed in Table~\ref{tbl:kinematics}. These stars may be a population of runaway stars, representing some of the first sub-solar hypervelocity stars found to date \citep{palladino:2014:7}. An in-depth study of these stars is beyond the scope of this work, however, a number of these stars will be discussed further in the kinematical study of high-velocity M dwarfs by Favia \& West (2014, in preparation). We will discuss scenarios that could explain an IR excess for a star moving with a high velocity relative to the ISM in \S\ref{discussion}. USNO-B/SDSS proper motions and our derived kinematics for all our stars are available in the online catalog.

\begin{deluxetable*}{clccccc}
\tabletypesize{\footnotesize}
\tablecolumns{7}
\tablewidth{0pt}
\tablecaption{High Velocity Candidate Kinematics\label{tbl:kinematics}}
\tablehead{
\colhead{Parameter} & \colhead{Reference} & \colhead{Candidate 27} &
\colhead{Candidate 50} & \colhead{Candidate 73} & \colhead{Candidate 106} & \colhead{Candidate 125} 
}
\startdata
\multirow{5}*{\minitab[c]{$\mu_\mathrm{R.A.}$ \\ (mas yr$^{-1}$)}} & \citet{munn:2004:3034,munn:2008:895} & $-349.5 \pm 17.1$ & $-24.6 \pm 3.4$ & $-185.1 \pm 3.0$ & $-65.9 \pm 3.4$ & $262.1 \pm 2.9$ \\
& \citet{monet:2003:984} & $-352 \pm 43$ & $-22 \pm 9$ & $-192\pm6$ & $-62 \pm 2$ & $266 \pm 4$\\
& \citet{lepine:2005:1483}\tablenotemark{a} & ... & ... & $-192 \pm 7$ & $-70 \pm 7$ & $235 \pm 7$\\
& \citet{roeser:2010:2440} & $-346.5 \pm 16.2$ & $-27.4 \pm 4.3$ & $-191.9\pm4.2$ & $-67.5 \pm 4.3$ & $266.3 \pm 3.7$\\
& \citet{salim:2003:1011} & ... & ... & $-189.4\pm5.5$ & ... & ... \vspace{0.1cm} \\ 
\hline \\ [-1.5ex]
\multirow{5}*{\minitab[c]{$\mu_\mathrm{Dec.}$ \\ (mas yr$^{-1}$)}} & \citet{munn:2004:3034,munn:2008:895} & $-212.2 \pm 17.1$ & $-87.6 \pm 3.4$ & $-159.8 \pm 3.0$ & $-145.5 \pm 3.4$ & $-85.4 \pm 2.9$ \\
& \citet{monet:2003:984} & $-230 \pm 39$ & $-80 \pm 7$ & $-152\pm4$ & $-140 \pm 3$ & $-80 \pm 3$ \\
& \citet{lepine:2005:1483}\tablenotemark{a} & ... & ... & $-170 \pm 7$ & $-150 \pm 7$ & $-87 \pm 7$ \\
& \citet{roeser:2010:2440} & $-236.6 \pm 16.2$ & $-91.1 \pm 4.3$ & $-160.9\pm4.2$ & $-150.8 \pm 4.3$ & $-87.1 \pm 3.7$ \\
& \citet{salim:2003:1011} & ... & ... & $-156.95\pm5.5$ & ... & ... \vspace{0.1cm} \\
\hline \\ [-1.5ex]
RV (km s$^{-1}$) && $-5.5 \pm 6.7$ & $97.1 \pm 8.9$ & $-50.7 \pm 4.6$ & $-51.1 \pm 5.0$ & $-129.0 \pm 2.4$ \\
$U$ (km s$^{-1}$) && $558.8 \pm 38.8$ & $-230.8 \pm 16.9$ & $-397.0 \pm 60.5$ & $357.5 \pm 57.6$ & $-246.7 \pm 24.3$ \\
$V$ (km s$^{-1}$) && $127.0 \pm 33.5$ & $-596.1 \pm 24.9$ & $-763.5 \pm 60.2$ & $-433.5 \pm 39.3$ & $-132.5 \pm 12.4$ \\
$W$ (km s$^{-1}$) && $-378.8 \pm 28.4$ & $-101.7 \pm 20.2$ & $-175.3 \pm 27.6$ & $216.5 \pm 21.3$ & $-328.0 \pm 28.1$ \\
$v_\mathrm{tot}$ (km s$^{-1}$) && $689.1 \pm 35.8$ & $640.5 \pm 23.9$ & $871.7 \pm 59.3$ & $597.9 \pm 45.2$ & $428.5 \pm 25.9$ 
\enddata
\tablenotetext{a}{Typical uncertainties stated.}
\end{deluxetable*}

\subsection{Metallicity}\label{metallicity}

	There is evidence that metallicity plays a vital role in the formation of stars and planets. Metallicity has been shown to correlate with the rate of giant planet occurrence \citep{gonzalez:1997:403, fischer:2005:1102, johnson:2010:905, mann:2012:90}. Studies of nearby clusters have shown that metal rich environments may allow primordial circumstellar disks to persist on longer timescales \citep{yasui:2009:54,yasui:2010:l113}. Additionally, accretion of planetary bodies may increase the metallicity of the host star \citep[e.g.,][]{farihi:2009:805}. Therefore, a detailed understanding of how metallicity affects disk evolution is crucial to understanding stellar and planetary system formation and evolution.

	It is challenging to measure the metal content of M dwarfs due to their large abundance of molecular features, many of which have incomplete line lists. Previous studies have attempted to find metallicity tracers using both photometric \citep{bonfils:2005:635, johnson:2009:933, schlaufman:2010:a105}, and spectroscopic methods \citep{lepine:2007:1235, rojas-ayala:2010:528, dhital:2012:67, mann:2012:90, onehag:2012:a33, rojas-ayala:2012:93, terrien:2012:l38, newton:2014:20}. Recently, \citet{mann:2013:52} completed an investigation of metallicity sensitive tracers in the optical and NIR spectra of late K and M dwarf companions to higher mass stars. The \citet{mann:2013:52} metallicity solutions are comprised of a set of equations based on equivalent widths and color indices. A number of metal sensitive features were found, including Na \textsc{i} (doublets at 8200 \AA\ and 2.208 \um), Ca \textsc{i} (1.616 \um\ and 1.621 \um), and K \textsc{i} (1.5176 \um), many of which were previously known to be metal sensitive. 
	
\begin{figure}
\plotone{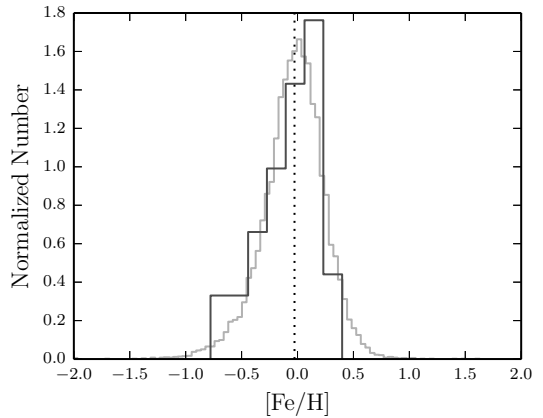}
\caption{The normalized distribution of $\metal$ measurements. We show stars from our combined sample (dark gray line) 
and the W11 catalog (light grey line) with $\sigma_\metal < 0.5$. Measurements were made using the method described in
\citet{mann:2013:52}. Only stars with a SpT $<$ dM7 were used. The median for the our sample is shown ($\metal \approx -0.03 \pm 0.28$; dotted line).
Our disk candidates appear to be distributed fairly evenly about solar metallicity, and appear to have a similar distribution to that of the W11
catalog. Dust caused from accretion of minor planets has been shown to increase the metal content of white dwarfs, however, no similar 
trend is seen here. \label{fig:metallicity}}
\end{figure}

	To study the metallicity of our sample, we used the SDSS spectra to compute equivalent widths and color indices in line with the methods defined in \citet{mann:2013:52}. To balance quality measurements with a statistically significant sample size, we used only stars with $\sigma_\metal < 0.5$ for our analysis, however, cutting on lower or higher values of $\sigma_\metal$, or removing the Orion candidates, did not significantly change our results. We find our sample to be distributed about solar metallicity, with a median value of $\metal \approx -0.03 \pm 0.28$, as shown in Figure~\ref{fig:metallicity}. This is similar to values computed for the W11 catalog ($\metal \approx -0.02 \pm 0.10$) using the same quality cut of $\sigma_\metal < 0.5$. For the scenario of tidal disruption and accretion of minor planets, we might expect our stars to be preferentially metal rich due to accretion, similar to results observed for white dwarfs with IR excesses \citep{jura:2003:l91,jura:2008:1785,farihi:2009:805}. However, for our stars we do not see a trend of high metallicity, making the tidal disruption and accretion scenario unlikely. Our stars do not have a significantly different metallicity distribution from the W11 catalog (as inferred by a K-S test; p-value $= 0.57$), therefore, we can draw no obvious conclusions about the metal content affecting, or being affected by, the mechanism for the observed IR excesses.

\subsection{Disk Fractions}\label{diskfrac}

	Our combined sample is large enough to study the nature of the excess IR flux in a Galactic context. Young star forming H\textsc{ii} regions are found most prominently in spiral arms, close to the Galactic plane. For example, \citet{lee:2008:1352} found that $\sim$90\% of embedded clusters are found within $\sim$160 pc of the Galactic plane. Therefore, we should expect young populations of stars to appear close to the Galactic plane with the number dropping off with vertical Galactic height. This ``Galactic stratigraphy" was used in W08, using magnetic activity as a proxy for youth. W08 found that the fraction of active M dwarfs declined with Galactic height, with different slopes for different spectral types, most likely due to the correlation between activity lifetime and spectral type. To test for a similar trend in our data, we performed a ``Galactic stratigraphy" analysis to investigate any age dependence for the observed IR excesses.
	
\begin{figure}
 \plotone{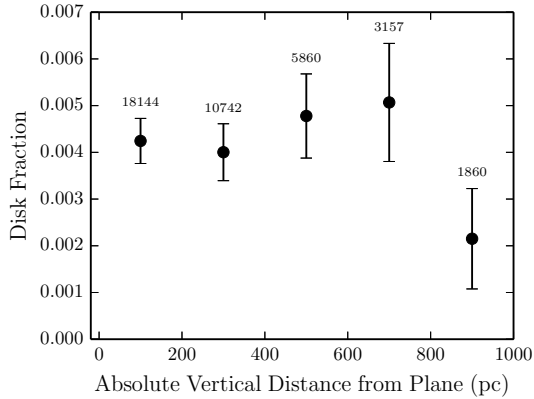}
\caption{Disk fractions (using $W3$ excess as a proxy) as a function of height using the combined sample without the Orion candidates. Error bars represent binomial errors. The number of stars in each bin are listed above their error bars. There is a slight declining trend seen at $\sim$700 pc, however, we do not see a strong height/age dependence. \label{fig:diskfrac1}}
\end{figure}

	Using our catalog of 41,120 stars for which we could have observed an IR excess (\S\ref{extragalactic}), we computed the fraction of stars with disks as a function of Galactic height. Figure~\ref{fig:diskfrac1} shows the fraction of disks (using $W3$ excess as a proxy), as a function of distance from the Galactic plane. Once the Orion candidates are removed, we see a fairly constant fraction until a distance of $\sim$700 pc, at which point we see a slight decline. Separating stars by spectral type does not change this result. As each Galactic height bin represents an ensemble of young and old stars, with younger stars more preferentially found closer to the plane, we surmise that the mechanism creating our observed dust populations happens randomly throughout low-mass stellar populations. We will explore possible mechanisms to create dust around old field stars in \S\ref{discussion}.

\subsection{Orion OB1 Candidate Stars}\label{orion}

\begin{figure*}
 \plotone{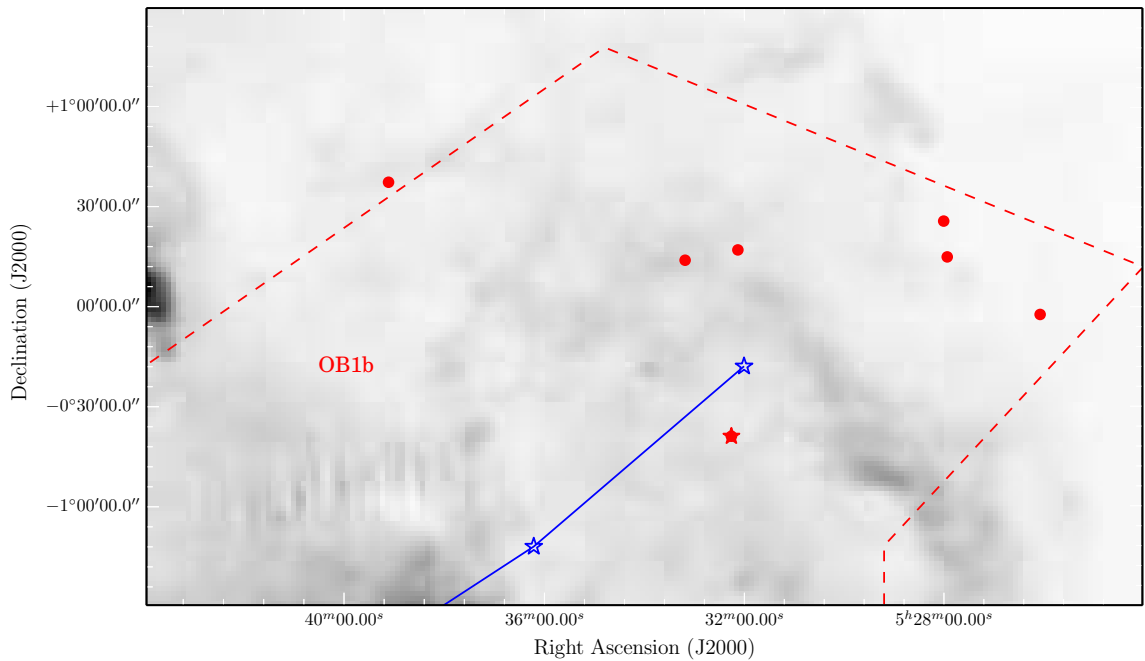}
\caption{Orion stars (red circles) overlaid on a cutout of the \textit{IRAS} IRIS 100 \um\ map \citep{miville-deschenes:2005:302}. The approximate boundary of the OB1b subassociation is shown \citep[red dashed line;][]{briceno:2007:1119}. Orion's Belt is shown for reference (blue line). Candidate 41 is shown as the red star.
\label{fig:orion}}
\end{figure*}

	Although the Orion complex and regions within the complex have been extensively researched \citep[see][and references therein]{bally:2008:459}, we found \orionsample\ stars that have not been previously linked to the region. All \orionsample\ candidates are found within a few degrees of Orion's Belt, a neighborhood with many young star-forming regions. Specifically, these stars are possible members of the Orion OB1 association. The positions of our Orion stars are shown in Figure~\ref{fig:orion}. Three of our stars fell into the Orion footprint investigated by \citet{caballero:2008:931}, but were not included in their sample, possibly due to their color selection criteria or depth of coverage. The majority of our Orion candidates have photometric distances between 100--340 pc, which distributes them approximately within the extent of the OB1 association \citep[$d \sim 100$--400 pc;][]{caballero:2007:903}. One of our Orion candidates may be foreground or background star, based on its kinematics (\S\ref{kinematics}). Basic information for the Orion candidates is listed in Table~\ref{tbl:orionparams}. Some of our Orion candidates showed a number of forbidden emission lines (e.g., [O \textsc{i}]), giving further evidence for their youth. One such star will be discussed below.

\begin{deluxetable*}{lcrrcccccc}
\tabletypesize{\scriptsize}
\tablecolumns{10}
\tablewidth{0pt}
\tablecaption{Orion Candidate Parameters\label{tbl:orionparams}}
\tablehead{
& \colhead{} & \colhead{R.A.} & \colhead{Decl.} & & & & & \\
\colhead{Catalog} & \colhead{\textit{WISE} } & \colhead{(J2000.0)} & \colhead{(J2000.0)} & \colhead{SpT\tablenotemark{a}} & \colhead{RV} & \colhead{$d$\tablenotemark{b}} & \colhead{Z} & \colhead{$A_V$\tablenotemark{c}} & \colhead{$R_V$\tablenotemark{d}} \\
\colhead{Number} & \colhead{Designation} & \colhead{(deg)} & \colhead{(deg)} & \colhead{$\pm 1$} & \colhead{(km s$^{-1}$)} & \colhead{(pc)} & \colhead{(pc)} &\colhead{(mags)} & 
}
\startdata
37.......... & J052604.21$-$000220.9 & 81.517565 & -0.039167 & M5 & 37.6 $\pm$ 4.7 & 130 & -27 & 0.41 $\pm$ 0.07 & 3.1 \\
38.......... & J052756.04+001459.5 & 81.983513 & 0.249883 & M4 & 28.2 $\pm$ 4.7 & 152 & -33 & 0.26 $\pm$ 0.05 & 5.0 \\
39.......... & J052800.30+002546.7 & 82.001268 & 0.429638 & M7 & 41.8 $\pm$ 6.2 & 122 & -23 & 0.00 $\pm$ 0.02 & 3.1 \\
40.......... & J053207.80+001704.9 & 83.032543 & 0.284724 & M2 & 32.2 $\pm$ 3.8 & 195 & -44 & 0.52 $\pm$ 0.02 & 3.1 \\
41.......... & J053215.59$-$003900.1 & 83.064987 & -0.650051 & M4 & 55.9 $\pm$ 5.5 & 241 & -59 & 0.21 $\pm$ 0.07 & 3.1 \\
42.......... & J053311.20+001358.5 & 83.296678 & 0.232939 & M3 & 30.8 $\pm$ 3.4 & 128 & -23 & 0.35 $\pm$ 0.06 & 3.1 \\
43.......... & J053906.74+003722.6 & 84.778096 & 0.622946 & M3 & 13.2 $\pm$ 2.6 & 331 & -75 & 0.48 $\pm$ 0.05 & 3.1 
\enddata
\tablenotetext{a}{Determined in \citet{west:2011:97}.}
\tablenotetext{b}{Photometric distances have typical uncertainties of $\sim$20\%.}
\tablenotetext{c}{Measured in \citet{jones:2011:44}.}
\tablenotetext{d}{Adopted parameters from \citet{jones:2011:44} measurements.}
\end{deluxetable*}

\subsubsection{Candidate 41}\label{ttauri}

\begin{figure*}
\plotone{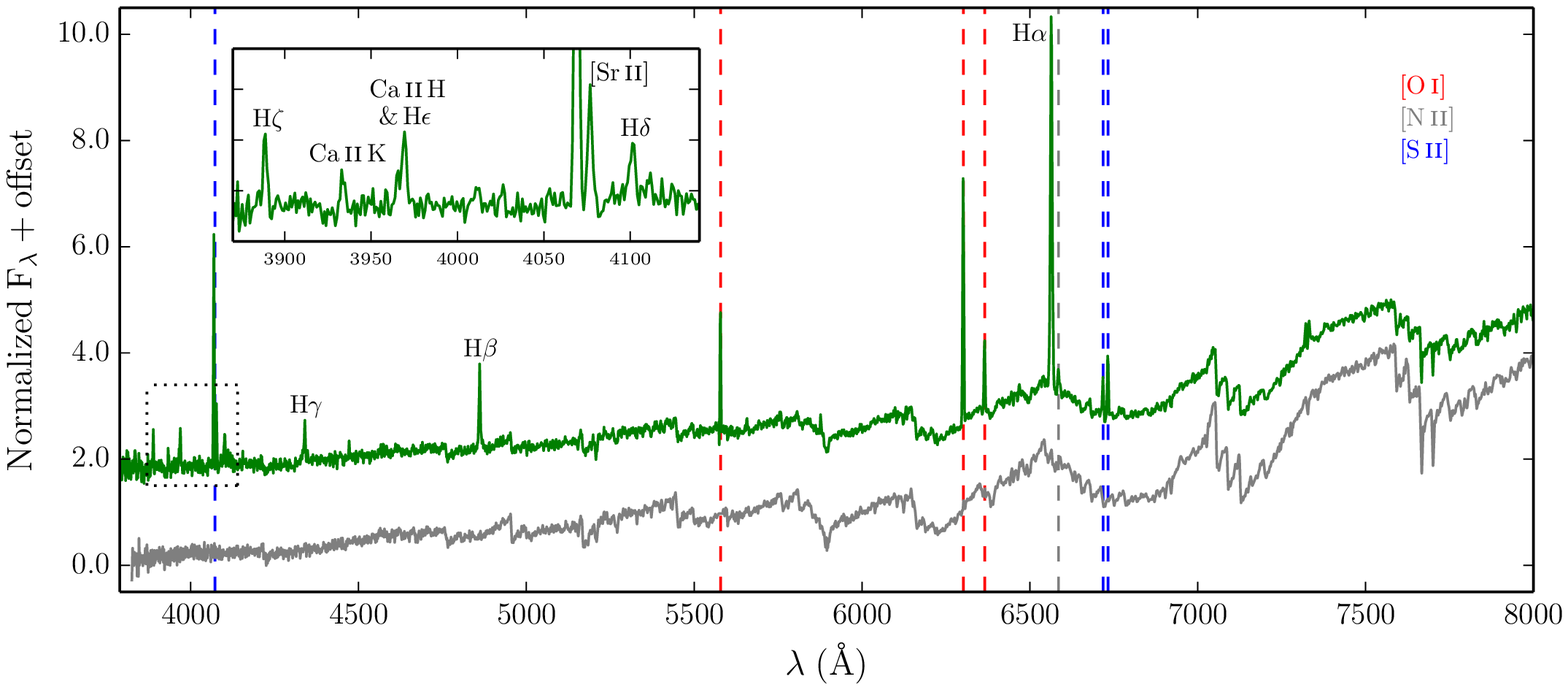}
\caption{SDSS spectrum of candidate 41 (green line) and a template spectrum of the same spectral type from \citet[][gray line]{bochanski:2007:531}.
There are a number of forbidden line transitions seen in this spectrum (line colors correspond to elements listed on the right side of the plot). 
These forbidden transitions are typically associated with T Tauri stars, and are an indication of accretion. 
The two emission lines $\sim$7300 \AA\ are most likely [O \textsc{ii}] lines \citep{kogure:2010:}.
The dotted box represents the area of the inset plot. Comparing the two spectra, veiling is apparent in our candidate spectrum. \label{fig:ttauri}}
\end{figure*}

	One of our candidates showed up as an outlier in our SED fits (see Fig.~\ref{fig:modelparams}). The SED fit for this star suffered from large uncertainties in both the radius and luminosity estimate, although the temperature fit was within the spread of its spectral type ($2900^{+600}_{-100}$ K; dM4). Closer inspection of this star uncovered strong Balmer emission lines as well as a number of forbidden lines within its spectrum ([O \textsc{i}], [N \textsc{ii}], and [S \textsc{ii}]), as shown in Figure~\ref{fig:ttauri}. Similar forbidden lines are observed in the spectrum of T Tauri stars \citep{appenzeller:1984:108}. Comparing the candidate spectrum to a template spectrum from \citet{bochanski:2007:531} shows veiling of the photosphere; whereby additional flux, typically from accretion, weakens spectral features. 

	This star is also one of our Orion OB1 candidates. The Orion OB1 association is assumed to have an age between $\sim$8--10 Myr \citep{briceno:2007:1119}. If we assume a temperature of 3,000 K for this star, using the isochrones from \citet{baraffe:1998:403, baraffe:2002:563} this star will have a radius of $\sim$$0.8 R_\odot$, the assumed radius from our parameter fits (\S\ref{methods}), at an age of $\sim$1--2 Myr. Even though there appears to be an age discrepancy, with the large uncertainties on the effective temperature and radius of this star, we cannot confidently estimate an age using isochrones. Candidate 41 is prime for follow-up observations to more accurately characterize its properties.
	
\section{Discussion}\label{discussion}

	We identified \wthreesample\ M dwarfs with IR excess at 12 and/or 22 \um, a typical signature of warm circumstellar material. From our youth analysis, we found that only \youthpercent\ show reliable youth diagnostics, and we conclude that the majority of stars in our sample have ages $\gtrsim 1$ Gyr; far older than the disk dispersal time for solar-type stars \citep[10s of Myr;][]{williams:2011:67}. This places the majority of our sample within the older field population. Unfortunately, without strict constraints on age we are not able to inform the timescale for disk dispersal in low-mass stars under the assumption we are observing primordial dust. Candidates in close proximity of the Orion OB1 association were more likely to show signs of lithium absorption, making them potentially younger than the rest of our sample ($\lesssim 100$ Myr). Due to their potential youth, these stars may still retain their primordial dust content. For the remaining stars, we examine possible scenarios to explain the appearance of warm dust.
	
	The minimum distances of the dust populations, typically found within the snow line of M dwarfs, and the high fractional luminosities imply that we may be viewing dust created by planetary impacts within the terrestrial zone, similar to conclusions drawn by \citet{weinberger:2011:72}. BD +20 307 is a spectroscopic binary composed of two F-type stars and is estimated to be older than 1 Gyr, with a large fractional IR luminosity \citep[$L_\mathrm{dust} / L_\ast \approx 0.032$;][]{weinberger:2011:72}. Using a measured \textit{Spitzer} spectrum, \citet{weinberger:2011:72} deduced that the most likely cause for a high fractional IR luminosity, and the crystallinity of their best-fit dust model, was a single giant impact of planetary bodies (specific incident kinetic energy of the two impacting objects $\gtrsim 10^5$ J km$^{-1}$). 
	
	A similar scenario has been used to explain IR excesses observed around older ($\gtrsim 1$ Gyr) FGK stars \citep{rhee:2008:777,moor:2009:l25,fujiwara:2010:l152,melis:2010:l57,weinberger:2011:72}. \citet{weinberger:2011:72} argue that a fifth terrestrial planet within a planetary system such as our own could remain stable for up to a Gyr. Simulations show that the planet's orbit could have taken it across the asteroid belt, perturbing asteroids into unstable orbits, increasing the impact rate of asteroids within the inner solar system, all prior to removal from our system \citep{chambers:2007:386}. This planet is also able to create collisions between terrestrial planets prior to leaving the system. This ``Planet V" scenario has been used as a possible explanation for the late heavy bombardment period within our Solar system \citep{chambers:2007:386}. Using numeric simulations \citet{laskar:2009:817} showed that over the course of a few Gyr, our own planetary system may become unstable and result in collisions of terrestrial planets, making this a possible scenario for terrestrial systems around M dwarfs. 
	
	In a NIR interferometric study of solar-type and higher-mass stars with debris disks undertaken by \citet{absil:2013:a104}, it was argued that collisions between large planetismals could create a significant, observable amount of dust. However, they deemed this an unlikely scenario for stars $\gtrsim1$ Gyr since these types of collisions are expected to happen during the final stages of planet formation \citep[$\lesssim$ a few 100s of Myr;][]{chambers:2004:241}. Clearly the timescales for such an occurrence still require further modeling to help constrain the timescale for collisions in planetary systems.
	
	Using statistics from \textit{Spitzer} and \textit{IRAS}, \citet{weinberger:2011:72} estimate that the giant impact rate for solar-type stars is $\gtrsim 0.2$ impacts per star during its main sequence lifetime. Following a similar argument, we can estimate the number of giant impacts for M dwarfs. The number of giant impacts per star can be estimated as $N_\mathrm{g}=f_\ast A L^{-1}$, where $f_\ast$ is the fraction of stars observed to have warm dust, $L$ is the lifetime of the collision products, and $A$ is the age of the stars surveyed. Assuming the total number of stars we could have observed with similar fractional IR luminosities within the W11 catalog to be stars with $J \leq 17$, we are left with 41,120 stars after our cuts (see \S\ref{diskfrac} for details). We observed \minusorion\ stars with warm dust populations that do not appear to be primordial in nature (we have omitted the Orion candidates), giving us a fraction of $f_\ast \approx 4.1\times10^{-3}$. \citet{weinberger:2011:72} used 100,000 years for $L$. This value was originally calculated by \citet{melis:2010:l57} for a collisional cascade started by planetary embryos with radii of $\sim$100 m, but was found to be similar to the 80,000 years found by \citet{weinberger:2011:72} for a collisional cascade started from a planet-sized impact.
	
	To estimate an average age for our observed stars, we used the activity lifetimes (i.e., the timescale during which H$\alpha$ should be observable in the stellar spectrum due to magnetic activity) results of \citet{west:2008:785}. For the 41,120 stars within the W11 catalog that made our cuts, we took the total number of inactive stars within a spectral type bin and multiplied those stars by the average activity lifetime for that spectral type bin from \citet{west:2008:785}. We did the same for active stars, assuming an age of half the activity lifetime for each spectral type bin. We then took the average age across the 41,120 stars, giving us an estimated average age of $A \approx 2.6$ Gyrs. We use 2.6 Gyrs as a lower limit on the age, as many of these fields stars are likely older than 5 Gyrs, which will only increase our collision estimate. Using our estimated quantities, we find the rate at which such planetary collisions occur around M dwarfs should be $\gtrsim 100$ impacts per star over its current age; three orders of magnitude larger than the number of impacts found by \citet{weinberger:2011:72} for A- through G-type stars.
	
	A high rate of giant impacts within the terrestrial planet zone for low-mass stars has repercussions for close-in planetary systems. The high occurrence rate of terrestrial planets orbiting low-mass stars \citep{dressing:2013:95,kopparapu:2013:l8} not only supports the possibility of this scenario, it also implies that a high rate of planetary collisions could significantly alter the habitability of such systems. N-body simulations for planet formation around low-mass stars also indicate that giant impacts could be quite common for these planetary systems \citep{ogihara:2009:824}. Future simulations should attempt to constrain the timescale and frequency for such impacts to occur, and the likelihood of observing such impacts within our Galaxy. Due to the assumed ages of our stars, we find giant impacts within the terrestrial zone to be the most plausible explanation for our observed IR excesses.
	
	To investigate whether our observed IR excesses could be due to \emph{primordial} dust requires an evolutionary study of the fraction of M dwarfs with IR excesses at different stellar ages. Unfortunately, detection limits and small sample sizes have hindered a thorough statistical analysis for disk lifetimes. We expect low-mass stars older than a few 10s of Myr to harbor cold disk populations ($T_\mathrm{dust} < 100$ K), rather than the warm disks in our study. The lack of cold dust populations found around M dwarfs in contrast to debris disks found around similar age, higher-mass stars indicates that primordial warm dust should not be found around low-mass stars $\gtrsim 1$ Gyr. However, this conclusion may be biased due to observational limits in detecting cold dust around low-mass stars. Future studies of \textit{Spitzer} and \textit{Herschel} data may further constrain the timescale for disk dispersal around M dwarfs. 
	
	If we consider the possibility that we are observing \emph{primordial} dust, we can explore kinematic arguments. The primary dispersal mechanism for low-mass stars is thought to be stellar wind drag \citep{plavchan:2005:1161,plavchan:2009:1068}, whereby the collisionally processed grains are removed by the proton wind rather than radiation pressure. If disk kinematics are able to overcome gravitational instabilities and stellar wind drag, a star may be able to retain its primordial dust content. Studies investigating planet masses around low-mass stars indicate that there is a paucity of Jupiter- versus sub-Neptune-sized planets \citep{howard:2012:15}. These findings have been attributed to low disk masses from which to build planets \citep{laughlin:2004:l73,kennedy:2008:502}. If there is a small enough reservoir of dust, a scenario in which the collision rate within low mass disks continues to reprocess the dust while inhibiting planet formation is a possibility for low-mass stars. Studies investigating asteroids as failed planets within our Solar System lend support to this argument \citep{coradini:2011:492,sierks:2011:487,russell:2012:684}. The likelihood of this scenario would need to be studied through dynamical simulations to explore its plausibility.
	
	For dust that is not primordial, we may be able to explain warm dust around older stellar populations through tidal disruption of planetary bodies. This is a phenomenon that has been observed around white dwarfs \citep[WDs; e.g.,][]{jura:2003:l91,jura:2008:1785,farihi:2009:805}. Through accretion of planetary bodies, these WDs show spectral signs of heavy elements in absorption \citep{holberg:1997:l127}. In the case of WDs, shedding of the outer layers during the transition from giant to dwarf is potentially responsible for perturbing the orbits of minor planetary bodies. No such mechanism exists for low-mass stars that have not evolved off the main sequence, making this scenario unlikely for our sample. Our estimated minimum orbital distances for our dust populations are also greater than the estimated Roche-limit for M dwarfs ($\lesssim 0.01$ AU), likely ruling out this scenario. Likewise, we can discount that our observed excesses are due to collisions fed by an asteroid belt, as dynamical models estimate much lower fractional IR luminosities than those observed \citep{wyatt:2007:569}.
	
	Another possible disruption scenario involves the inherently chaotic nature of planetary orbits \citep{murray:1999:1877}. For close-in giant planets with unstable orbits that degrade to come within the corotation radius of their host star, it is expected that they would lose angular momentum quickly to the host star through tidal interactions. This would result in tidal disruption once the planet comes within the host star's Roche limit \citep{schlaufman:2013:143}. As the material is accreted onto the host star the dust would heat up and appear as an IR excess. The dynamical arguments of \citet{schlaufman:2013:143} did not include low-mass stars or Earth-sized planets, however, if we assume their arguments scale to lower-masses we find that the corotation radius for a $0.5M_\odot$ star is $\lesssim 0.1$ AU. Mass scaling relations for the core accretion process responsible for forming giant planets, estimate the distance of the snow-line for low-mass stars during the time of giant planet formation is $> 1$ AU \citep{laughlin:2004:l73,adams:2005:913,ida:2005:1045,kennedy:2008:502}. Due to the relative paucity of Jupiter-sized planets around low-mass stars \citep{johnson:2007:833}, we find it unlikely that such a migration scenario occurs at high frequency. A study investigating the likelihood of this scenario for Earth-sized planets around low-mass stars should be undertaken in the future.
	
	To provide a separate explanation for IR excesses observed for our high-velocity stars, we consider $\zeta$ Oph, a high-mass, runaway star with a shock clearly visible in the \textit{WISE} image archives. This star shows significant IR flux in the form of thermal emission from the bowshock of its astrosphere. Crude estimates using aperture photometry on the \textit{WISE} images give a flux ratio of $F_{\rm star+bowshock}/F_{\rm star} > 10$ for both the $W3$ and $W4$ bands. This phenomenon has been observed around a number of runaway massive stars \citep{peri:2012:a108}, but requires further investigation to determine if it is a viable scenario for high-velocity, low-mass stars.

\section{Summary}\label{summary}

	Using the \citet{west:2011:97} SDSS catalog of 70,841 spectroscopically classified M dwarfs, we searched for IR excesses as signatures of warm circumstellar disks using \textit{WISE} data. Applying a number of selection criteria, including our own SDSS/{\em WISE} color-color selection criteria using $r-z$ color as a temperature proxy, we discovered \totalsample\ stars with significant excess flux at 12 and/or 22 \um, most probably due to circumstellar material. We explored the possibility that any of our excesses could be due to an ultracool companion and found that none of our candidates exhibited flux levels representative of an ultracool companion at $\geq 3\sigma$ level. Using statistical arguments, we concluded that our observed IR excesses arising from chance alignment with extragalactic sources or Galactic cirrus contamination were unlikely.
	 
	 Fitting our observational data to SEDs, we modeled both the stellar and dust components of our fits. We find that our fractional IR luminosities, $L_\mathrm{dust} / L_\ast$, are high ($> 10^{-3}$). These values are similar to those found around higher-mass stars showing IR excesses at ages $\gtrsim 1$ Gyr \citep{weinberger:2011:72}. Dust equilibrium models put our disks typically within a few tenths of an AU to their host star, and possibly closer depending upon the parameters chosen. None of our dust populations have a minimum orbital distance $< 0.01$ AU, the approximate Roche limit for low-mass stars, indicating that tidal disruption of planetary bodies is not a likely cause of the observed IR excess. Our estimated dust masses are slightly smaller than masses found in debris disks around higher-mass stars \citep[$\sim$$10^{-3}$--$10^{-4} M_\mathrm{moon}$;][]{chen:2005:1372}, and can be attributed to lower-mass disks around lower-mass stars. 
	 	 
	 We tested our sample for youth by observing lithium absorption, measuring surface gravity tracers (CaH 3, K \textsc{i}, and Na \textsc{i}) and magnetic activity tracers (H$\alpha$, UV flux). Only a fraction of the stars in our sample (\youthpercent) show multiple signs of youth, and we conclude that the majority of our stars are field dwarfs older than a Gyr. Seven of our candidate stars lie in the footprint of the Orion OB1 association. Most of the Orion candidate stars are magnetically active ($\sim$86\%), show low surface gravity relative to the W11 catalog of M dwarfs, or show lithium absorption, and in some cases a combination of several youth tracers. Stars showing lithium absorption likely indicate an age $\lesssim 100$ Myr. Although previously unidentified, most of these stars are likely members of the OB1 association. 
	 
	 An analysis of $UVW$ space motions uncovered the Orion sample to be a kinematically younger population of stars relative to our entire sample. The remainder of our candidates exhibited kinematics similar to typical disk populations, with a few high-velocity exceptions. Five stars exhibited total velocities greater than 400 km s$^{-1}$ relative to the local standard of rest. Neither the dynamical origin of these stars, nor the mechanism to propel them to such high-velocities are known. It is possible that if these stars are moving supersonically through dense regions of the ISM, that a significant bowshock could produce high levels of IR flux, supplanting the need for a circumstellar disk to explain an IR excess. 
	 
	 To further investigate the timescale of this phenomenon, we employed the use of ``Galactic Stratigraphy," using 12 \um\ excess as a disk (and hence age) proxy. We find that the disk fraction for stars as a function of vertical height from the Galactic plane remains fairly constant out to a height of $\sim$700 pc at which point the fraction begins to drop off. We do not see a significant difference for early versus late spectral types. This likely implies there is no strong age dependence of the mechanism responsible for this phenomenon.
	 
	 The estimated ages of our stars ($\gtrsim 1$ Gyr) imply that we are not observing primordial dust content. Instead, we find the most likely explanation for our IR excesses is giant impacts of terrestrial planetary bodies. This hypothesis is supported by observations of similar high-fractional IR luminosity around Solar-type stars with ages $\gtrsim 1$ Gyr \citep{weinberger:2011:72}. Also, the lack of observed disks around low-mass stars with ages greater than a few 10s of Myr suggests that warm primordial dust content does not persist around low-mass stars for longer than 1 Gyr. The estimated number of giant impacts per low-mass star over its main sequence lifetime, $N_\mathrm{g} \gtrsim 100$, indicates that such occurrences may be relatively common around low-mass stars. By enlarging our sample size with photometric M dwarfs from the latest SDSS data release \citep[DR10;][]{ahn:2014:17}, we should identify thousands more such excesses around field M dwarfs. These stars will also prove important targets for the Atacama Large Millimeter Array (ALMA) and future IR missions such as the \textit{James Webb Space Telescope} and the \textit{Space Infrared Telescope for Cosmology and Astrophysics}.

\acknowledgments

	The authors would first like to thank the anonymous referees for their extremely helpful comments which greatly improved the quality of this study. The authors would also like to thank Dan Feldman, Catherine Espaillat, Dylan Morgan, and Jan-Marie Andersen for reading this manuscript and providing helpful comments, and Saurav Dhital, Isabella Trierweiler, Lara Rangel, Chris Boyter, Dan Clemens, Jim Davenport, Kevin Covey, Alycia Weinberger, Sarah Schmidt, Tom Megeath, Christine Chen, Keivan Stassun, Jonathan Foster, Debra Fischer, Meredith Hughes, and Mark Zastrow for their helpful discussions. C.A.T. would like to acknowledge the Ford Foundation for financial support. C.A.T. thanks Gibor Basri, Geoff Marcy, and Jeff Valenti for making their radial velocity fitting code available. A.A.W acknowledges funding from NSF grants AST-1109273 and AST-1255568. A.A.W. and C.A.T. also acknowledge the support of the Research Corporation for Science Advancement's Cottrell Scholarship.

	Funding for the SDSS and SDSS-II has been provided by the Alfred P. Sloan Foundation, the Participating Institutions, the National Science Foundation, the U.S. Department of Energy, the National Aeronautics and Space Administration, the Japanese Monbukagakusho, the Max Planck Society, and the Higher Education Funding Council for England. The SDSS Web Site is http://www.sdss.org/. The SDSS is managed by the Astrophysical Research Consortium for the Participating Institutions. The Participating Institutions are the American Museum of Natural History, Astrophysical Institute Potsdam, University of Basel, University of Cambridge, Case Western Reserve University, University of Chicago, Drexel University, Fermilab, the Institute for Advanced Study, the Japan Participation Group, Johns Hopkins University, the Joint Institute for Nuclear Astrophysics, the Kavli Institute for Particle Astrophysics and Cosmology, the Korean Scientist Group, the Chinese Academy of Sciences (LAMOST), Los Alamos National Laboratory, the Max-Planck-Institute for Astronomy (MPIA), the Max-Planck-Institute for Astrophysics (MPA), New Mexico State University, Ohio State University, University of Pittsburgh, University of Portsmouth, Princeton University, the United States Naval Observatory, and the University of Washington.
 
	This publication makes use of data products from the Two Micron All Sky Survey, which is a joint project of the University of Massachusetts and the Infrared Processing and Analysis Center/California Institute of Technology, funded by the National Aeronautics and Space Administration and the National Science Foundation. This publication also makes use of data products from the \textit{Wide-field Infrared Survey Explorer}, which is a joint project of the University of California, Los Angeles, and the Jet Propulsion Laboratory/California Institute of Technology, funded by the National Aeronautics and Space Administration. 
		
	This research made use of Montage, funded by the National Aeronautics and Space Administration's Earth Science Technology Office, Computation Technologies Project, under Cooperative Agreement Number NCC5-626 between NASA and the California Institute of Technology. Montage is maintained by the NASA/IPAC Infrared Science Archive. This research also made use of Astropy, a community-developed core Python package for Astronomy \citep{astropy-collaboration:2013:a33}. Figures in this work were created using the Python based graphics environment Matplotlib \citep{hunter:2007:90}.

\bibliography{arxiv}
\bibliographystyle{apj}

\end{document}